\documentclass[aps,prb,reprint,amsmath,amssymb,floatfix,superscriptaddress]{revtex4-2}

\usepackage{color,graphicx,multirow,scrextend,hyperref}

\usepackage[ignoreunlbld,norefs,nocites]{refcheck}

\begin{document}

\title{Quantum Monte Carlo study of three-dimensional Coulomb
  complexes: trions and biexcitons; hydrogen molecules and ions;
  helium hydride cations; and positronic and muonic complexes}

\author{F.\ Marsusi} \email{marsusi@aut.ac.ir} \affiliation{Department
  of Energy Engineering and Physics, Amirkabir University of
  Technology, PO Box 15875-4413, Tehran, Iran}

\author{E.\ Mostaani} \affiliation{Cambridge Graphene Centre,
  University of Cambridge, 9 J.\ J.\ Thomson Avenue, Cambridge, CB3
  0FA, United Kingdom}

\author{N.\ D.\ Drummond} \affiliation{Department of Physics,
  Lancaster University, Lancaster LA1 4YB, United Kingdom}

\date{\today}

\begin{abstract}
Three-dimensional (3D) excitonic complexes influence the
optoelectronic properties of bulk semiconductors. More generally,
correlated few-particle molecules and ions, held together by pairwise
Coulomb potentials, play a fundamental role in a variety of fields in
physics and chemistry. Based on statistically exact diffusion quantum
Monte Carlo calculations, we have studied excitonic three- and
four-body complexes (trions and biexcitons) in bulk 3D semiconductors
as well as a range of small molecules and ions in which the nuclei are
treated as quantum particles on an equal footing with the
electrons. We present interpolation formulas that predict the binding
energies of these complexes either in bulk semiconductors or in free
space. By evaluating pair distribution functions within quantum Monte
Carlo simulations, we examine the importance of harmonic and
anharmonic vibrational effects in small molecules.
\end{abstract}

\maketitle


\section{Introduction}

Excitons (X) are hydrogen-like bound states of excited electron-hole
pairs in semiconductors. They significantly affect the optical
properties of direct-gap semiconductors, especially at low
temperature, giving rise to narrow peaks below the conduction band
edge in optical spectra. Excitons are electrically neutral composite
bosons. If an exciton binds to a free electron or hole in a
semiconductor, a negatively or positively charged trion (X$^\pm$) is
formed, which can be regarded as an exotic analog of a hydride H$^-$
anion or a dihydrogen H$_2^+$ cation. Trion formation generally
requires an imbalance in the populations of electrons and holes, e.g.,
when photoexcitation takes place in a doped semiconductor. Trion
binding energies are much smaller than exciton binding energies
because trions are held together by a charge-induced dipole
interaction rather than a charge-opposite charge Coulomb
attraction. In analogy to dihydrogen H$_2$ or dipositronium Ps$_2$
molecules, a pair of excitons may form a bound state called a
biexciton (X$_2$). Biexciton formation does not require an imbalance
in the populations of electrons and holes.

Although ionic cores modify the electronic dispersion in a
semiconductor enormously, around the band edges the dispersion is
free-particle-like (i.e., quadratic) and hence the effects of ionic
cores can often be described by an effective mass approximation.
Furthermore, ions and core electrons screen the electron and hole
charges, making the Coulomb potential between charge carriers in a
semiconductor much weaker than the Coulomb interactions in an isolated
atomic or molecular system. Especially in crystals of cubic symmetry,
the screened Coulomb potential can often be described by an isotropic,
homogeneous, static permittivity. Since this permittivity is generally
large in covalent semiconductors, electrons and holes bind weakly to
form so-called Mott-Wannier excitons. These excitons are weakly
localized, have a radius larger than the lattice constant, and do not
significantly alter the atomic structure. They therefore act as free
complexes moving within the semiconductor. In many bulk
three-dimensional (3D) semiconductors, the trion and biexciton lines
in optical spectra cannot easily be identified due to their weak
intensity and proximity to the exciton peak \cite{Narita_1976,
  Taniguchi_1976}; therefore, theoretical investigations, including
both analytical and numerical methods, are essential in this field.

The Mott-Wannier Hamiltonian that describes charge-carrier complexes
in crystals with isotropic effective masses and permittivities is of
the same form as the nonrelativistic Hamiltonian for few-particle
atoms, molecules, or ions in free space \cite{Kezerashvili_2019}. In
this work we study small numbers of interacting charges, including
trions and biexcitons in semiconductors as well as isolated ions and
molecules of hydrogen, and positronic and muonic species. We focus on
the ground-state properties of each complex, for which the spatial
wave function is nodeless and the particles can be treated as
distinguishable. Thus, for example, we examine para-H$_2$ (in which
the electrons and protons are in spin singlet configurations) and
ortho-D$_2$ (in which the electrons are in a spin singlet but the
deuterons are in an ortho spin configuration).

Different analytical many-body formalisms have previously been applied
to compute 3D trion and biexciton total energies using the Coulomb
potential. Shiau \textit{et al.}\, solved approximately the
Schr\"{o}dinger equations of trions and biexcitons, using a free
exciton basis \cite{Shiau_2012,Shiau_2013}. They treated
exciton-electron and exciton-exciton interactions within the composite
boson many-body formalism.  However, the predicted binding energies
show a discrepancy with previous results obtained using variational
trial wave functions \cite{Thilagam_1997,Sergeev_2001} due to a
restriction of the exciton basis to the low-lying $s$-like excitonic
wave function. In another work, Combescot calculated the binding
energy of a trion from a general exact solution to the three-body
problem based on the scattering $T$-matrix \cite{Combescot_2017}. The
numerical results obtained in Ref.\ \onlinecite{Combescot_2017} are in
good agreement with the variational energies reported in
Refs.\ \onlinecite{Brinkman_1973}, \onlinecite{Wehner_1969}, and
\onlinecite{Korobov_2000}, and were used to propose a formula for
calculating 3D trion binding energies. Despite the straightforwardness
and accuracy of the analytical method employed in
Ref.\ \onlinecite{Combescot_2017}, it cannot easily be used to study
larger complexes such as biexcitons.  We present a series of numerical
results obtained using the variational and diffusion quantum Monte
Carlo (VMC and DMC) approaches \cite{Foulkes_2001} to predict the
ground-state binding energies of three- and four-body excitonic
complexes in 3D\@. We use trial wave functions consisting of pairing
functions multiplied by Jastrow correlation factors. Since the ground
states of the complexes that we study are formed of distinguishable
particles, the corresponding wave functions are nodeless. This is an
important point because the DMC method gives the ground-state energy
of such a system without bias (in the limit of zero time step,
adequate equilibration, and infinite walker population); there is no
fixed-node error.  The VMC and DMC methods have previously been used
to study 3D trions and biexcitons
\cite{Bressanini_1997,Tsuchiya_2000b} and the para-H$_2$ molecule
\cite{Hunt_2018}. Bressanini \textit{et al.}\ presented total and
binding energies of exotic four-particle Coulomb complexes consisting
of two oppositely charged heavy particles and two oppositely charged
light particles using VMC and DMC methods \cite{Bressanini_1997}.
Here we focus on the far more commonly encountered case where
particles of the same charge also have the same mass.

Throughout, we assume isotropic electron and hole masses and
permittivities, and so our model is appropriate for cubic-symmetry
direct-gap III-V binary semiconductors, which are crucially important
in electronic research and technology due to their high carrier
mobilities, tunable gaps, and the availability of well established
growth and characterization techniques. We have fitted algebraic
functions to our DMC binding energies such that the fractional error
in the fitted binding energy at each data point is less than 0.01\%.
Using these interpolation formulas we are able to predict the binding
energies of Mott-Wannier trions and biexcitons in bulk semiconductors
as functions of the electron and hole effective masses $m_\text{e}$
and $m_\text{h}$ and the static permittivity $\epsilon$. We have also
examined extreme electron-hole mass ratios
$\sigma=m_\text{e}/m_\text{h}$ and we present an analysis of limiting
behavior near $\sigma=0$ and $\sigma=\infty$, which is of relevance to
real three- and four-body systems such as the Ps$^-$, Mu$^-$, H$^-$,
D$^-$, T$^-$, Mu$_2^+$, Ps$_2^+$, H$_2^+$, D$_2^+$, T$_2^+$, Ps$_2$,
Mu$_2$, H$_2$, D$_2$, and T$_2$ ions and molecules that are at the
heart of thermonuclear processes, astronomy, and atomic, molecular,
and chemical physics. We define all these complexes in terms of their
constituents in Tables \ref{table:trion_energies},
\ref{table:biex_energies}, and \ref{table:misc_complexes}.  In the
heavy-``hole'' limit $\sigma \to 0$, the electron and hole degrees of
freedom decouple and we simply need to solve the Schr\"{o}dinger
equation for the electrons in the presence of fixed positive
particles; the resulting electronic ground-state energy as a function
of the hole separation provides a Born-Oppenheimer (BO) potential
energy surface within which the holes move. By fitting interpolation
formulas to the DMC energy against hole-hole distance, we are able to
calculate the static equilibrium distance and some important
spectroscopic constants \cite{Dunham_1932}. Furthermore, without
making the BO approximation, we have investigated physical properties
of dihydrogen molecules and cations such as the dynamical
(nonadiabatic) mean nucleus-nucleus distances, which can be deduced
from the pair distribution functions (PDFs)\@. These PDFs provide
valuable information about the spatial size of an excitonic
complex. Furthermore, the PDFs allow the evaluation of the
electron-hole contact density, which is an important factor in the
recombination rate. Finally, for completeness, we present quantum
Monte Carlo (QMC) data for other small Coulomb complexes of physical
importance: mixed-isotope hydrogen molecules, helium hydride cations,
and positronic and muonic hydrogen molecules.

\section{Computational methodology and details}

\subsection{Hamiltonian for excitonic complexes}

We model Mott-Wannier excitonic complexes in a 3D semiconductor within
the isotropic effective mass approximation. The Coulomb interactions
between the electrons and holes are isotropically screened by the
static permittivity $\epsilon$ of the crystal. For a system consisting
of $N_\text{e}$ electrons and $N_\text{h}$ holes, the Hamiltonian is
\begin{widetext}
\begin{equation} \hat{H} = -\sum_{i=1}^{N_\text{e}}
\frac{\hbar^2\nabla_{\text{e},i}^2}{2m_\text{e}} -
\sum_{i=1}^{N_\text{h}}
\frac{\hbar^2\nabla_{\text{h},i}^2}{2m_\text{h}} -
\sum_{i=1}^{N_\text{e}} \sum_{j=1}^{N_\text{h}}
\frac{e^2}{4\pi\epsilon|{\bf r}_{\text{e},i}-{\bf r}_{\text{h},j}|} +
\sum_{i=1}^{N_\text{e}-1} \sum_{j=i+1}^{N_\text{e}} \frac{e^2}{4\pi
  \epsilon|{\bf r}_{\text{e},i}-{\bf r}_{\text{e},j}|} +
\sum_{i=1}^{N_\text{h}-1} \sum_{j=i+1}^{N_\text{h}} \frac{e^2}{4\pi
  \epsilon|{\bf r}_{\text{h},i}-{\bf r}_{\text{h},j}|}, \end{equation}
where $m_\text{e}$ and $m_\text{h}$ denote the electron and hole
effective masses, respectively, and ${\bf r}_{\text{e},i}$ and ${\bf
  r}_{\text{h},j}$ are the position vectors of the $i$th electron and
$j$th hole, respectively. We introduce dimensionless positions ${\bf
  r}'={\bf r}/a_0^*$, such that $\nabla'=a_0^*\nabla$ and
${(\nabla')}^2={(a_0^*)}^2\nabla^2$, where
$a_0^*=4\pi\epsilon\hbar^2/(\mu e^2)$ is the exciton Bohr radius and
$\mu=m_\text{e}m_\text{h}/(m_\text{e}+m_\text{h})$ is the
electron-hole reduced mass, and dimensionless masses
$m_\text{e}'=m_\text{e}/\mu$ and $m_\text{h}'=m_\text{h}/\mu$. The
Hamiltonian can then be written as
\begin{eqnarray} \hat{H} & = & \frac{\mu e^4}{{(4\pi\epsilon)}^2\hbar^2} \Bigg[
 -\frac{1}{2} \sum_{i=1}^{N_\text{e}}
 \frac{{(\nabla'_{\text{e},i})}^2}{m_\text{e}'} - \frac{1}{2}
 \sum_{i=1}^{N_\text{h}}
 \frac{{(\nabla'_{\text{h},i})}^2}{m_\text{h}'} \nonumber \\ & &
 \qquad \qquad \qquad {} - \sum_{i=1}^{N_\text{e}}
 \sum_{j=1}^{N_\text{h}} \frac{1}{|{\bf r}'_{\text{e},i}-{\bf
     r}'_{\text{h},j}|} + \sum_{i=1}^{N_\text{e}-1}
 \sum_{j=i+1}^{N_\text{e}} \frac{1}{|{\bf r}'_{\text{e},i}-{\bf
     r}'_{\text{e},j}|} + \sum_{i=1}^{N_\text{h}-1}
 \sum_{j=i+1}^{N_\text{h}} \frac{1}{|{\bf r}'_{\text{h},i}-{\bf
     r}'_{\text{h},j}|}
 \Bigg]. \label{eq:Hamiltonian_eu} \end{eqnarray}
\end{widetext}
The constant $\mu e^4/[{(4\pi\epsilon)}^2\hbar^2]$ is our unit of
energy, the exciton Hartree. Henceforth we will drop the primes from
the nondimensional lengths and masses, and we will use
``e.u.''\ (excitonic units) to indicate that a length is in units of
the exciton Bohr radius, or that an energy is in units of the exciton
Hartree, or that mass is in units of the electron-hole reduced mass.
In the limit $m_\text{e} \ll m_\text{h}$, we have $\mu=m_\text{e}$ and
hence e.u.\ reduce to Hartree atomic units (a.u.).

For excitonic complexes with $N_\text{e}\geq 2$ or $N_\text{h} \geq 2$
the Hamiltonian of Eq.\ (\ref{eq:Hamiltonian_eu}) is
$\sigma$-dependent. For $N_\text{e}=N_\text{h}=1$, we rewrite
Eq.\ (\ref{eq:Hamiltonian_eu}) in terms of the center of mass position
${\bf r}_\mu=(m_\text{e}{\bf r}_\text{e}+m_\text{h}{\bf
  r}_\text{h})/(m_\text{e}+m_\text{h})$ and the position of the
electron relative to the hole ${\bf r}={\bf r}_\text{e}-{\bf
  r}_\text{h}$, giving (in e.u.)
\begin{equation} \hat{H}=\hat{H}_\text{r} + \hat{H}_\mu = \left(
 -\frac{1}{2}\nabla^2-\frac{1}{|{\bf r}|} \right) - \frac{\mu}{2M}
 \nabla_\mu^2, \end{equation} with $M=m_\text{e}+m_\text{h}$.
$\hat{H}_\text{r}$ is the Hamiltonian term describing the internal
motion of the system, due to the interaction between the electrons and
the holes, and $\hat{H}_\mu$ describes the kinetic energy of the
center of mass, which is zero in the ground state. Since the
$\hat{H}_\text{r}$ and $\hat{H}_\mu$ terms are independent, the wave
function can be written as a product $\Psi({\bf r},{\bf
  r}_\mu)=\phi_\text{r}({\bf r})\phi_\mu({\bf r}_\mu)$, and the
exciton energy can be found from the first part
\begin{equation} \left(-\frac{1}{2}\nabla^2-\frac{1}{|{\bf r}|}
\right)\phi_\text{r}({\bf r})=E_\text{X}\phi_\text{r}({\bf
  r}). \label{eq:X_SE} \end{equation} Equation (\ref{eq:X_SE}) is
$\sigma$-independent.

\subsection{QMC calculations: excitonic wave functions and Jastrow terms}

We calculated the total energies of complexes by solving the few-body
Schr\"{o}dinger equation using the VMC and DMC methods
\cite{Foulkes_2001}. We employed trial wave functions of the form
$\Psi=\exp(J)\Psi_\text{S}$. The $\Psi_\text{S}$ part of the wave
function is a sum of products of excitonic pairing orbitals:
\begin{eqnarray} \Psi_\text{S} & = & \phi_1(r_{\text{e},1;\text{h},1})
\phi_1(r_{\text{e},2;\text{h},2}) \phi_2(r_{\text{e},1;\text{h},2})
\phi_2(r_{\text{e},2;\text{h},1}) \nonumber \\ & & {} +
\phi_2(r_{\text{e},1;\text{h},1}) \phi_2(r_{\text{e},2;\text{h},2})
\phi_1(r_{\text{e},1;\text{h},2}) \phi_1(r_{\text{e},2;\text{h},1}),
\nonumber \\ \end{eqnarray} where $r_{\text{e},i;\text{h},j}=|{\bf
  r}_{\text{e},i}-{\bf r}_{\text{h},j}|$, for biexcitons
\cite{Bauer_2013} and
\begin{equation} \Psi_\text{S} = \phi_1(r_{\text{e},1;\text{h},1})
\phi_2(r_{\text{e},2;\text{h},1}) + \phi_2(r_{\text{e},1;\text{h},1})
\phi_1(r_{\text{e},2;\text{h},1}) \end{equation} for negative trions.
The pairing orbitals are of form
\begin{equation} \phi_i(r)= \exp\left(-r^2/[a_i(b_i+r)]\right), \end{equation}
where $\{a_i\}$ and $\{b_i\}$ are optimizable parameters. The pairing
orbitals only couple electron-hole pairs, but their long-range
exponential behavior binds the complex. The pairing orbitals do not
enforce the Kato cusp conditions; instead these are enforced via the
Jastrow factor $\exp(J)$. The Jastrow exponent $J$ consists of
two-body polynomial ($U$) and three-body polynomial ($H$) terms that
are truncated at finite range over a few exciton Bohr radii
\cite{Drummond_2004}. In simulations with fixed holes (i.e.,
$\sigma=0$), particle-ion ($\chi$) and particle-particle-ion ($F$)
terms were also included in $J$. Free parameters in the wave function
were optimized within VMC by minimizing the energy variance
\cite{Umrigar_1988,Drummond_2005} and then energy expectation value
\cite{Umrigar_2007}.

In the trial wave functions of complexes with very small but nonzero
$\sigma$, we have included an additional two-body Jastrow term of the
form $-c{(r-r_0)}^2$ between the heavy holes, where $c$ and $r_0$ are
optimizable parameters. This term violates both the short-range (Kato
cusp) behavior and the long-range exponential behavior, but is
appropriate for particle pairs whose motion is primarily vibrational.
Including this term lowers the VMC total energies of the molecular
hydrogen isotopes H$_2$, D$_2$, and T$_2$ by $7.3(3) \times 10^{-4}$,
$7.10(2) \times 10^{-3}$, and $6.00(3) \times 10^{-3}$ e.u.,
respectively.

To ensure that time step bias is negligible, we have examined the
effect of varying the time step on the DMC total energy for a positive
trion and a biexciton at a very small mass ratio $\sigma$; if time
step bias is negligible at an extreme mass ratio then it is certain to
be negligible at mass ratios closer to 1. In
Figs.\ \ref{fig:app_trion_time_step} and
\ref{fig:app_biexciton_time_step} of Appendix \ref{app:time_step}, we
compare the zero-time-step DMC energy obtained from a linear fit to
two small DMC time steps (0.01 and 0.0025 e.u.)\ with the
zero-time-step DMC energy obtained from a quadratic fit to data at six
time steps over a wider range. The extrapolated energies are in
statistical agreement. Hence we performed our production DMC
calculations using the two small time steps in the ratio 1:4, with the
target walker population being varied in inverse proportion to the
time step, and we linearly extrapolated the resulting DMC energies to
zero time step and therefore infinite population. Since we study
nodeless ground state wave functions, the fixed-node DMC energy is
exact; nevertheless, it is still desirable to obtain an accurate trial
wave function to improve the statistical efficiency of the algorithm
and to improve the expectation values of operators such as the PDF
that do not commute with the Hamiltonian.

All our VMC and DMC calculations were performed using the
\textsc{casino} code \cite{Needs_2020}.

\subsection{PDFs and contact interactions between charge carriers}

Although the Mott-Wannier excitons in 3D crystals extend over many
unit cells, there is a nonzero probability density that the charge
carriers are found at the same point in the crystal. In this case,
significant local exchange and correlation effects are expected. This
implies an additional, perturbative, pairwise contact interaction
potential \cite{Mostaani_2017,Feenberg_2012}
\begin{eqnarray} \hat{V}_\text{contact} & = & A^\text{ee} \sum_{i=1}^{N_\text{e}-1}
\sum_{j=i+1}^{N_\text{e}} \delta({\bf r}_{\text{e},i}-{\bf
  r}_{\text{e},j}) \nonumber \\ & & {} + A^\text{hh}
\sum_{i=1}^{N_\text{h}-1} \sum_{j=i+1}^{N_\text{h}} \delta({\bf
  r}_{\text{h},i}-{\bf r}_{\text{h},j}) \nonumber \\ & & {} +
A^\text{eh} \sum_{i=1}^{N_\text{e}} \sum_{j=1}^{N_\text{h}}
\delta({\bf r}_{\text{e},i}-{\bf r}_{\text{h},j}), \end{eqnarray}
where $A^\text{ee}$, $A^\text{hh}$, and $A^\text{eh}$ are constants
and can be found via \textit{ab initio} calculations or by fitting to
experimental results. The first-order perturbative expectation value
of the contact interaction potential is $\langle
\hat{V}_\text{contact}\rangle = A^\text{ee} g_\text{ee}({\bf 0}) +
A^\text{hh} g_\text{hh}({\bf 0}) + A^\text{eh} g_\text{eh}({\bf 0})$
where the electron-electron, hole-hole, and electron-hole PDFs are
\begin{eqnarray} g_\text{ee}({\bf r}) & = & \left< \sum_{i=1}^{N_\text{e}-1}
\sum_{j=i+1}^{N_\text{e}} \delta({\bf r}-({\bf r}_{\text{e},i}-{\bf
  r}_{\text{e},j})) \right> \\ g_\text{hh}({\bf r}) & = & \left<
\sum_{i=1}^{N_\text{h}-1} \sum_{j=i+1}^{N_\text{h}} \delta({\bf
  r}-({\bf r}_{\text{h},i}-{\bf r}_{\text{h},j})) \right>
\\ g_\text{eh}({\bf r}) & = & \left< \sum_{i=1}^{N_\text{e}}
\sum_{j=1}^{N_\text{h}} \delta({\bf r}-({\bf r}_{\text{e},i}-{\bf
  r}_{\text{h},j})) \right>, \end{eqnarray} respectively. In addition
to perturbative corrections due to contact interactions, the PDF gives
important information about an excitonic complex. The recombination
rate of an excitonic complex is proportional to the electron-hole
contact PDF \cite{Mostaani_2017}. Furthermore, the spatial size and
shape of a charge-carrier complex can be found directly from the
PDF\@.

The errors in the VMC and DMC estimates of each PDF depend linearly on
the error in the trial wave function. However, the error in the
extrapolated estimate (twice the DMC estimate minus the VMC estimate)
is quadratic in the error in the trial wave function
\cite{Ceperley_1979}. Here, we report extrapolated PDFs.

\section{Numerical results}

\subsection{Excitons}

Equation (\ref{eq:X_SE}) is of the form of the Schr\"{o}dinger
equation for a hydrogen atom and its well-known solution results in an
energy spectrum for bound states
\begin{equation} E_\text{X}=-\frac{1}{2n^2},
\end{equation}
where $n=1,2,\ldots$. In particular, the exciton ground-state energy
is $E_\text{X}=-1/2$ e.u., independent of the electron-hole mass
ratio.  To convert to ``real'' units (as opposed to e.u.)\ we need the
permittivity $\epsilon$ and electron and hole effective masses to
calculate the exciton Hartree as $\mu
e^4/[{(4\pi\epsilon)}^2\hbar^2]$.

The band effective mass approximation is valid if the ground state
energy of the exciton is much smaller than the corresponding
semiconductor energy gap \cite{Henry_1978}. This condition leads to a
very large exciton Bohr radius $a_0^*$, such that the exciton extends
over many crystal sites. Under such conditions, a crystal can
accurately be treated as a continuous medium, and electrons and holes
are well described by the effective mass approximation with statically
screened Coulomb interactions between charge carriers.

Equations (\ref{eq:Hamiltonian_eu}) and (\ref{eq:X_SE}) are obtained
under the assumption of isotropic electron and hole effective masses
and isotropic permittivities. Such a model is most suitable for cubic
zincblende-structure semiconductors such as InAs, InP, GaAs, InN, and
InSb. For these direct-gap binary III-V compounds, the lowest
conduction band minimum is isotropic, and it occurs at the Brillouin
zone (BZ) center. The degenerate valence band maxima corresponding to
heavy and light holes in these compounds also occur at the BZ
center. The hole masses are anisotropic in binary III-V compounds
\cite{luque2011handbook,Adachi_1993}; nevertheless, isotropic
approximations to the hole mass are widely used. Heavy holes lead to
larger exciton binding energies and hence we take the hole mass to be
that of heavy holes. In the diamond-structure group-IV semiconductors
Si and Ge, the conduction band minima are away from the BZ center (on
the $\Gamma$--X line) and are therefore strongly anisotropic, with
ellipsoidal symmetry. This anisotropy imposes great complexity on the
excitons in Si and Ge crystals. However, by noting the large exciton
Bohr radii in these crystals (see Table \ref{table:material_params})
and the cubic symmetry of the lattice, which results in a diagonal
effective mass tensor, one can use the simple spherical optical mass
average approximation to calculate the exciton ground state energy
\cite{Henry_1978}. The optical average electron mass $m_\text{e}$ is
defined via $m_\text{e}^{-1}=(2m_\text{et}^{-1}+m_\text{el}^{-1})/3$,
where $m_\text{et}$ and $m_\text{el}$ are the electron masses in the
transverse and longitudinal directions, respectively. In Si and Ge the
valence band is anisotropic around its maximum at $\Gamma$; we
therefore used the spherically averaged heavy hole effective mass in
our calculations.

\begin{table}[!htbp]
\centering
\caption{Electron effective mass $m_\text{e}$ in III-V semiconductors
  or optical average effective mass $m_\text{e}$ in group-IV
  semiconductors, spherically averaged hole effective mass
  $m_\text{h}$, static relative permittivity $\epsilon_\text{r}$,
  exciton Bohr radius $a_0^*$, exciton ground-state total energy
  $E_\text{X}$ (from the hydrogenic model and compared with available
  experimental data). \label{table:material_params}}
\begin{tabular}{lccccccc}
\hline \hline

\multirow{2}{*}{Crystal} & $m_\text{e}$ & $m_\text{h}$ &
\multirow{2}{*}{$\epsilon_\text{r}$} & \multicolumn{2}{c}{$a_0^*$
  ({\AA})} & \multicolumn{2}{c}{$E_\text{X}$ (meV)} \\

& (a.u.) & (a.u.) & & Theo. & Exp. & Theo. & Exp. \\

\hline

Si & $0.26$\footnote{Parameters taken from
Ref.\ \onlinecite{Levinshtein_1997}. \label{fn:handbook_si}} &
$0.49$\footref{fn:handbook_si} & $11.7$\footref{fn:handbook_si} &
$36.14$ & $44.3$\footnote{Data taken from
Ref.\ \onlinecite{Lipari_1971}. \label{fn:opav_Lipari}} & $-17.18$ &
$-14.7$\footref{fn:opav_Lipari} \\

Ge & $0.12$\footref{fn:handbook_si} & $0.33$\footref{fn:handbook_si} &
$16.2$ \footref{fn:handbook_si}& $97.42$ &
$127$\footref{fn:opav_Lipari} & $-4.56$ &
$-2.1$\footref{fn:opav_Lipari} \\

\hline

GaAs\footnote{Parameters taken from Ref.\ \onlinecite{Filikhin_2018}.}
& $0.067$ & $0.51$ & $12.90$ & $115.27$ & & $-4.84$
& \begin{tabular}{@{}c@{}} $-4.1(1)$\footnote{Data taken from
    Refs.\ \onlinecite{Alperovich_1976} and \onlinecite{Moore_1996}.}
    \\ $-4.8$\footnote{Data taken from
    Ref.\ \onlinecite{White_1972}.} \end{tabular}
\\ InAs\footnote{Parameters taken from
Ref.\ \onlinecite{Schubert_2006}.} & $0.022$ & $0.40$ & $15.1$ &
$383.18$ & & $-1.24$ & $-1$\footnote{Data taken from
Ref.\ \onlinecite{Tang_1997}.} \\

InSb\footnote{Parameters taken from
Ref.\ \onlinecite{Bouarissa_1999}.}  & $0.03$ & $0.41$ & $16.8$ &
$318.02$ & & $-1.35$ & \\

InP\footnote{Parameters taken from Ref.\ \onlinecite{Kim_2009}.} &
$0.089$ & $0.414$ & $12.09$\footref{dis:Turner} & $87.34$ & & $-6.82$
& $-4.0$\footnote{Data taken from
Ref.\ \onlinecite{Turner_1964}. \label{dis:Turner}} \\

\hline \hline
\end{tabular}
\end{table}

The physical parameters that we require to calculate the binding
energies of excitons in a selection of important semiconductors are
given in Table \ref{table:material_params}.  For semiconductors such
as Ge or InSb, even at low temperatures of 100 K (or equivalently 8.6
meV), thermal fluctuations may overcome the small exciton binding
energy. Consequently, exciton lines in photospectra can only be
detected only below $\sim 100$ K\@.

The exciton energy discussed in this section will be used to calculate
the binding energies of trions and biexcitons in subsequent sections.

\subsection{Trions}

\subsubsection{Binding energies}

The energy difference between the exciton peak and the peak of a
larger excitonic complex in a photoluminescence experiment is equal to
the energy required to separate a single exciton from the complex. For
freely moving trions and biexcitons, this is also the energy required
to break the complex into its most energetically favorable daughter
products, so we refer to this energy difference as the binding energy
of the complex.

\begin{figure}[!htbp]
\centering \includegraphics[clip,width=1\linewidth]{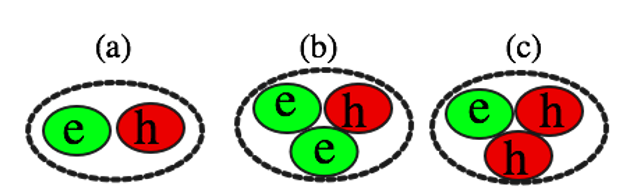}
\caption{(a) An exciton X is a bound electron-hole pair. (b) A
  negative trion is a negatively charged exciton X$^-$. (c) A positive
  trion is a positively charged exciton
  X$^+$. \label{fig:X_T_diagrams}}
\end{figure}

A positive trion (X$^+$) is a positively charged complex consisting of
two distinguishable holes and a single electron. Likewise, a negative
trion (X$^-$) is a negatively charged complex consisting of two
distinguishable electrons and one hole: see
Fig.\ \ref{fig:X_T_diagrams}. According to the above definition, the
binding energy of a positive or negative trion is the energy required
to separate the trion into a bound exciton and a free hole or
electron, respectively:
\begin{equation} E_{\text{X}^\pm}^\text{B}=E_\text{X}-E_{\text{X}^\pm},
\end{equation}
where $E_{\text{X}^\pm}$ is the ground state total energy of the trion
and $E_\text{X}$ is the ground state energy of the exciton. Note that
the ground-state energy of the free hole or electron is zero.

Generally, the hole effective mass in semiconductors is larger than
the electron effective mass because valence bands show weaker energy
dispersion, and often the electron-hole mass ratio $\sigma$ lies in
the range $0.01 < \sigma < 1$. Since the Coulomb interaction is
symmetric in terms of particle charge, the binding energy of a
negative trion with electron-hole mass ratio $\sigma$ is equal to the
binding energy of a positive trion with electron-hole mass ratio
$1/\sigma$. In fact, it is often more convenient to show the binding
energy of a trion in terms of the rescaled mass ratio
$x=\sigma/(1+\sigma) \in [0,1]$ rather than $\sigma \in [0,\infty)$.
  The binding energy of a negative trion with rescaled mass ratio $x$
  is equal to the binding energy of a positive trion with rescaled
  mass ratio $1-x$.

DMC binding energies of negative and positive trions at different mass
ratios $\sigma$  are listed  in Appendix  \ref{app:raw_data}, together
with  previous theoretical  data where  available. The  binding energy
reaches its maximum  value for two heavy holes and  one light electron
($\sigma=0$) or equivalently,  two heavy electrons and  one light hole
($\sigma\to\infty$).

\begin{figure}[!htbp]
\centering \includegraphics[clip,width=1\linewidth]{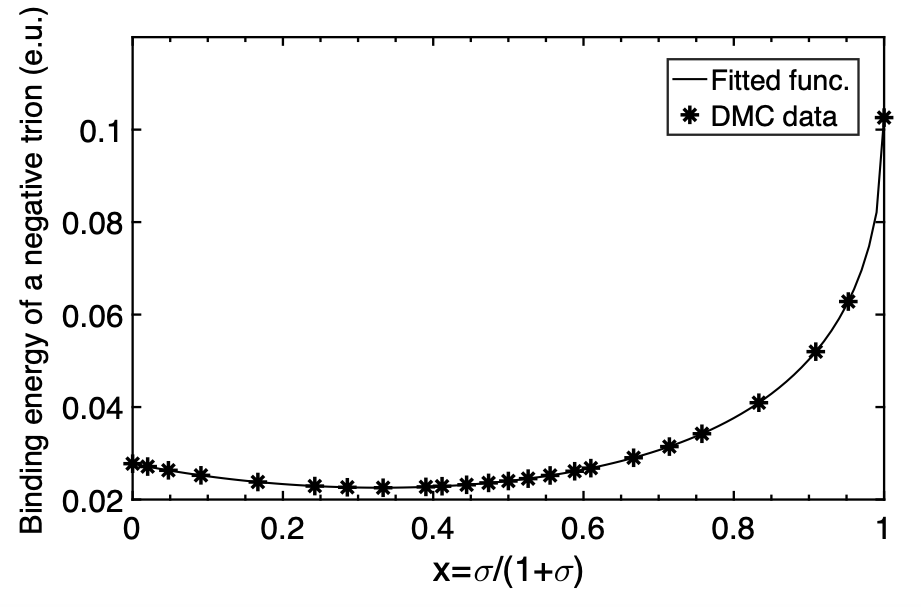}
\caption{DMC binding energies of negative trions against rescaled mass
  ratio $x=\sigma/(1+\sigma)$. The solid line shows the fit of
  Eq.\ (\ref{eq:trion_fit}). \label{fig:be_vs_x}}
\end{figure}

In the case $\sigma \to 0$ ($m_\text{e} \ll m_\text{h}$), we have an
X$^+$ consisting of a light electron moving in the field of two slowly
moving heavy holes, and the BO approximation can be applied to the
Hamiltonian of Eq.\ (\ref{eq:Hamiltonian_eu}) to separate the
electron's contribution to the total energy from the holes'
contributions. Let $R_\text{eq}$ be the position of the minimum of the
BO potential energy between two heavy holes; then the BO potential
energy at distance $r$ near $R_\text{eq}$ can be expanded as a Taylor
series:
\begin{eqnarray} U_\text{hh}(r) & = & U_\text{hh}(R_\text{eq})
+ \frac{1}{2} U_\text{hh}''(R_\text{eq}) {(r-R_\text{eq})}^2 +
\dots. \nonumber \\ & \equiv & U_\text{hh}(R_\text{eq})+\frac{1}{2}
\mu_\text{hh} \omega^2{(r-R_\text{eq})}^2. \end{eqnarray} The
hole-hole reduced mass in e.u.\ is $\mu_\text{hh}=m_\text{h}/2 =
(1+1/\sigma)/2=1/(2x)$.  Hence the vibrational frequency $\omega$ is
\begin{equation} \omega = \sqrt{2U_\text{hh}''(R_\text{eq})\,x}.
\label{eq:freq} \end{equation}
From Eq.\ (\ref{eq:freq}), the ground state energy of an X$^+$ in the
harmonic approximation is
\begin{equation} E \approx U_\text{hh}(R_\text{eq}) + \frac{1}{2} \omega
= U_\text{hh}(R_\text{eq}) + \frac{1}{2}
\sqrt{{2U_\text{hh}''(R_\text{eq})x}}. \label{eq:harmonic_approx}
\end{equation}
Thus the energy and hence binding energy of a positive trion increases
as $\sqrt{x}$ at small $x$ (and by charge symmetry, the energy of a
negative trion goes as $\sqrt{1-x}$ near $x=1$). Equation
(\ref{eq:harmonic_approx}) suggests that a suitable fitting function
for the binding energy of a negative trion is a Pad\'{e} function in
powers of $\sqrt{1-x}$:
\begin{equation}
E_{\text{X}^-}^\text{B} = \frac{\sum_{i=0}^3 a_i
  {(1-x)}^{i/2}}{1+\sum_{j=1}^3 b_j {(1-x)}^{j/2}},
\label{eq:trion_fit}
\end{equation}
where the values of the fitting parameters $\{a_i\}$ and $\{b_j\}$ are
presented in Table \ref{table:trion_fit_params}. In
Fig.\ \ref{fig:be_vs_x} the formula in Eq.\ (\ref{eq:trion_fit}) is
plotted against $x$ and compared with the original DMC data. The
summed square of deviations (SSE) from the DMC data is $8.37 \times
10^{-10}$ e.u., with a root-mean-square error (RMSE) of $7.128 \times
10^{-6}$ e.u. Equation (\ref{eq:trion_fit}) also describes positive
trions, provided $1-x$ is replaced by $x$.

\begin{table}[!htbp]
\centering
\caption{Fitted parameters in Eq.\ (\ref{eq:trion_fit}) for negative
  trion binding energies. \label{table:trion_fit_params}}
\begin{tabular}{lc}
\hline \hline

Parameter & Value (e.u.) \\

\hline

$a_0$ & $0.10259977858492200$ \\

$a_1$ & $-0.19032032332604387$ \\

$a_2$ & $0.12107317507399042$ \\

$a_3$ & $-0.013040190842318337$ \\

$b_1$ & $0.34760392556048214$ \\

$b_2$ & $-0.20820698951970068$ \\

$b_3$ & $-0.40782706820695203$ \\

\hline \hline
\end{tabular}
\end{table}

Equation (\ref{eq:trion_fit}) enables us to predict the binding energy
of a trion in a semiconductor in units of exciton Hartree given the
electron-hole mass ratio. If we also know the actual electron and hole
masses and permittivity, we can evaluate the exciton Hartree and hence
find the binding energy in real units. In Table \ref{table:trion_BEs}
we present examples of predicted binding energies of negative and
positive trions using Eq.\ (\ref{eq:trion_fit}) for a range of
technologically important semiconductors. In each case, the positive
trion forms a stronger bound state than the negative trion. This
appears to contradict the finding of Ref.\ \cite{Filikhin_2018},
obtained by solving the Faddeev equations within the effective mass
approximation, that a positive trion is unbound for semiconductors
such as GaAs. In each case, DMC predicts the trion binding energy to
be an order of magnitude smaller than the corresponding exciton
binding energy. Therefore, one expects the exciton peak to hide the
trion peak in each case.  Also, biexciton peaks may be difficult to
observe in experiments because the sub-meV binding energies are often
less than experimental uncertainties. In the following sections, we
will demonstrate another important application of our study in the
field of atomic and molecular physics.

\begin{table}[!htbp]
\centering
\caption{Predicted binding energies of negative and positive trions
  (X$^-$) and biexcitons (X$_2$) in bulk semiconductors, using
  Eqs.\ (\ref{eq:trion_fit}) and (\ref{eq:biex_fit}), respectively, to
  interpolate our DMC data. The material parameters are shown in Table
  \ref{table:material_params}, while the fitting parameters in the
  interpolation formulas are given in Tables
  \ref{table:trion_fit_params} and \ref{table:biex_fit_params}.
 \label{table:trion_BEs}}
\begin{tabular}{lccc}
\hline \hline

Crystal & $E_{\text{X}^-}^\text{B}$ (meV) & $E_{\text{X}^+}^\text{B}$
(meV) & $E_{\text{X}_2}^\text{B}$ (meV) \\

\hline

\multirow{2}{*}{GaAs} & $ 0.23910(3)$ &  $0.46066(4)$ & $0.67338(2)$
\\ & $0.5$\footnote{Data taken from Ref.\ \onlinecite{Filikhin_2018}.}
& \\

InAs & $ 0.065198(6)$ & $0.15279(1)$ & $0.238160(6)$ \\

InSb & $0.069477(6)$ & $0.15356(1)$ & $0.235254(6)$ \\

InP & $0.32245(4)$ & $0.54338(5)$ & $0.75443(3)$ \\

Si & $0.77530(9)$ & $ 0.97733(9)$ & $1.24781(9)$ \\

Ge & $0.20753(2)$ & $0.29744(3)$& $0.38983(3)$ \\

\hline \hline
\end{tabular}
\end{table}

\subsubsection{Mass effects in anions and molecular cations of hydrogen
 isotopes}

Molecular cations of hydrogen isotopes (i.e., H$_2^+$, D$_2^+$, and
T$_2^+$) are effectively trions consisting of two extremely heavy
``holes'' (the nuclei) and one light electron. The binding energy of
such a cation is the energy required to dissociate the system into a
neutral atom and a single nucleus. On the other hand, atomic anions
such as H$^-$, D$^-$, and T$^-$ can be viewed as negative trions
consisting of an extremely heavy ``hole'' and two light electrons. For
these anions, the binding energy is the energy needed to separate an
electron to infinite distance from the neutral atom and is equal to
the electron affinity of the neutral atom.

We performed DMC simulations of these ions and we report the resulting
total energies and binding energies in Table
\ref{table:trion_energies}. We take the proton mass, deuteron mass,
and triton mass to be $m_{\text{p}^+}=1836.152673440001$ a.u.,
$m_{\text{d}^+}=3670.482967853717$ a.u., and
$m_{\text{t}^+}=5496.921535729647$ a.u., respectively
\cite{CODATA_2018}.  The energy required to separate an electron from
a hydrogen anion slightly increases with isotope mass, as expected
from Fig.\ \ref{fig:be_vs_x}, and the corresponding energy approaches
the $\sigma=0$ limit of a negatively charged trion. The same trend is
observed for Ps$^-$ and Mu$^-$ (a bound state of an antimuon and two
electrons), while the calculated values are in good agreement with the
available experimental data.

\begin{table*}[!htbp]   
\centering
\caption{Ground-state total energy and binding energy of various ions
  obtained by DMC and compared with previous (experimental) works,
  where possible.  d$^+$ and t$^+$ denote a deuteron and a triton,
  respectively. 
\label{table:trion_energies}}
\begin{tabular}{lcccccc}
\hline \hline

 \multirow{2}{*}{Ion} & \multicolumn{3}{c}{Total energy (e.u.)} &
 \multicolumn{3}{c}{Binding energy (eV)} \\

 & DMC & Eq.\ (\ref{eq:trion_fit}) & Prev.\ works & DMC &
 Eq.\ (\ref{eq:trion_fit}) & Exp. \\

 \hline

  Ps$^-$ (e$^-$e$^-$e$^+$) & $-0.52401(1)$ & $-0.524009(2)$ &
  $0.524010$\footnote{Data taken from Ref.\ \onlinecite{Drake_2005}.}
  & $0.3267(1)$ & $0.32666(3)$ & $0.33$\footnote{Data taken from
  Ref.\ \onlinecite{Nagashima_2008}.}  \\

  Mu$^-$ (e$^-$e$^-\mu^+$) & $-0.52759(5)$ & $-0.527607(3)$ & &
  $0.747(1)$ & $0.74761(8)$ & $0.75$\footnote{Data taken from
  Ref.\ \onlinecite{Dudnikov_2018}.} \\

 H$^-$ (e$^-$e$^-$p$^+$) & $-0.52762(7)$ & $-0.527748(4)$ & &
 $0.751(2)$ & $0.7547(1)$ & $0.754$\footnote{Data taken from
 Ref.\ \onlinecite{Lykke_1991}. \label{fn:be_Lykke}} \\

D$^-$ (e$^-$e$^-$d$^+$) & $-0.527742(4)$ & $-0.527757(4)$ & &
$0.7547(1)$ & $0.7551(1)$ & $0.755$\footref{fn:be_Lykke} \\

 T$^-$ (e$^-$e$^-$t$^+$) & $-0.527744(3)$ & $-0.527760(4)$ & &
 $0.75482(8)$ & $0.7553(1)$ & \\

 Mu$_2^+$ (e$^-\mu^+\mu^+$) & $-0.587951(2)$ & $-0.58795(2)$ & &
 $2.38175(5)$ & $2.3817(5)$ \\

 H$_2^+$ (e$^-$p$^+$p$^+$) & $-0.5974636(7)$ & $-0.59745(1)$ &
 \begin{tabular}{@{}c@{}} $-0.60263461$\footnote{Data taken from
  Ref.\ \onlinecite{Schaad_1970}.}
   \\ $-0.597464275221235$\footnote{Data taken from
   Ref.\ \cite{Li_2007}} \end{tabular} & $2.65068(2)$ & $2.6503(3)$ &
 \begin{tabular}{@{}c@{}} $2.651$\footnote{Data taken from Ref.\
 \onlinecite{Mills_2007}. \label{fn:Mills}} \end{tabular} \\

 D$_2^+$ (e$^-$d$^+$d$^+$) & $-0.59893(3)$ & $-0.59893(1)$ & &
 $2.6913(8)$ & $2.6913(3)$ & $2.691$\footref{fn:Mills} \\

 T$_2^+$ (e$^-$t$^+$t$^+$) & $-0.59956(5)$ & $-0.59959(1)$ & &
 $2.709(1)$ & $2.7095(3)$ & \\

 X$^+$ (e$^-$h$^+$h$^+$), with $\sigma=0$ & $-0.60265(2)$ &
 $-0.60260(1)$ & $-0.6025$\footnote{Data taken from
 Ref.\ \onlinecite{Usukura_1999}.} & $2.7933(5)$ & $2.79189(3)$& \\

 \hline \hline
\end{tabular}
\end{table*}

Similarly, DMC predicts a slight increase in binding energy with mass
ratio for the three molecular cations of hydrogen isotopes, as seen in
Table \ref{table:trion_energies}. The corresponding values approach
the $\sigma=0$ limit for a positively charged trion, so that the
heavier isotope T$_2^+$ has a 3\% larger binding energy than the
lighter isotope H$_2^+$.

For comparison, we also show the total energy and binding energy of
the hydrogen ions calculated using Eq.\ (\ref{eq:trion_fit}) in Table
\ref{table:trion_energies}. There is excellent agreement between the
DMC results and Eq.\ (\ref{eq:trion_fit}). Consequently, knowing the
electron-hole mass ratio, Eq.\ (\ref{eq:trion_fit}) can be used to
predict the total energy and binding energy of either a trion in an
isotropic semiconductor or a Coulomb complex in free space.

\subsubsection{BO potential energy curve}

\begin{figure}[!htbp] 
\centering \includegraphics[clip,width=1\linewidth]{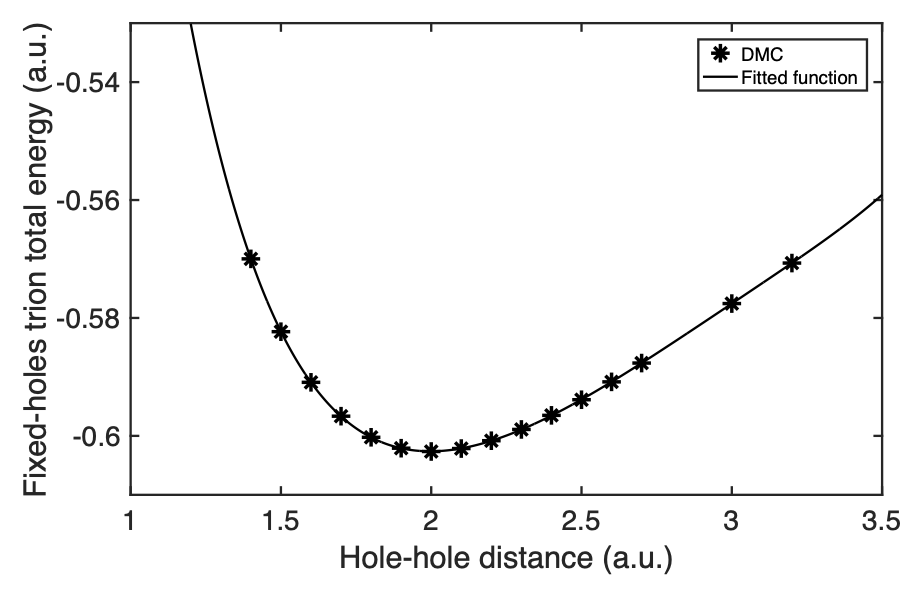}
\caption{DMC BO potential energy, i.e., DMC total energy of a positive
  trion in the infinite mass limit of two heavy holes and one light
  electron ($\sigma=0$) against the hole-hole distance. The solid
  curve shows a fit of Eq.\ (\ref{eq:BO_fit}) to the DMC
  data. \label{fig:bo_trion}}
\end{figure}

The extreme electron-hole mass ratio of $\sigma=0$ is very important
in atomic and molecular physics. The resulting BO potential allows us
to compute important spectroscopic data. We calculated the BO
potential energy curve as a function of hole-hole separation by
solving the Schr\"{o}dinger equation for a single electron in the
presence of two fixed holes using DMC\@. We selected a wide range of
hole-hole distances, between 1.4 and 3.5 a.u.  DMC total energies for
each hole-hole separation are given in Appendix
\ref{app:dmc_bo_data}. The interaction between nuclei in a diatomic
molecule or ion is often described by a Morse potential. The Morse
potential can qualitatively describe the BO potential at very short
and very large separations, as shown in Appendix
\ref{app:bo_fits}. However, because the Morse potential shows a
significant deviation from the DMC data near the equilibrium
separation, the resulting spectroscopic data based on the Morse model
disagree with experiments. Instead, we found that a degree-six
polynomial 
\begin{equation} U_\text{hh}(r) = \sum_{i=0}^6 p_{i} r^i, \label{eq:BO_fit}
\end{equation}
fits our DMC data well near the equilibrium separation, as seen in
Fig.\ \ref{fig:bo_trion}. The SSE and RMSE of the fit are $1.061
\times 10^{-8}$ and $3.434 \times 10^{-5}$ a.u, respectively, and the
maximum fractional deviation of the fitted function from the DMC data
is less than 0.007\%. The fitted coefficients $\{p_i\}$ are given in
Appendix \ref{app:dmc_bo_data}.

For infinite hole mass, the electron-hole reduced mass is
$\mu=m_\text{e}$, and assuming that $\epsilon_\text{r}=\epsilon_0$
(i.e., the complex is in free space), the exciton Bohr radius and
exciton Hartree energy are equal to the atomic Bohr radius and Hartree
energy (1 e.u.${}=1$ a.u.).

From Eq.\ (\ref{eq:BO_fit}) the minimum energy of a positive charged
trion in the fixed hole limit occurs at $R_\text{eq}=1.9970(5)$ e.u.,
which agrees well with the prediction of Schaad \textit{et al.}, who
used Burrau's method to separate the Schr\"{o}dinger equation for the
single electron in H$_2^+$ in confocal elliptical coordinates
\cite{Schaad_1970,erikson1949note}: see Table \ref{table:trion_rdf_spec}.

\begin{figure}[!htbp]
\centering
\includegraphics[clip,width=1\linewidth]{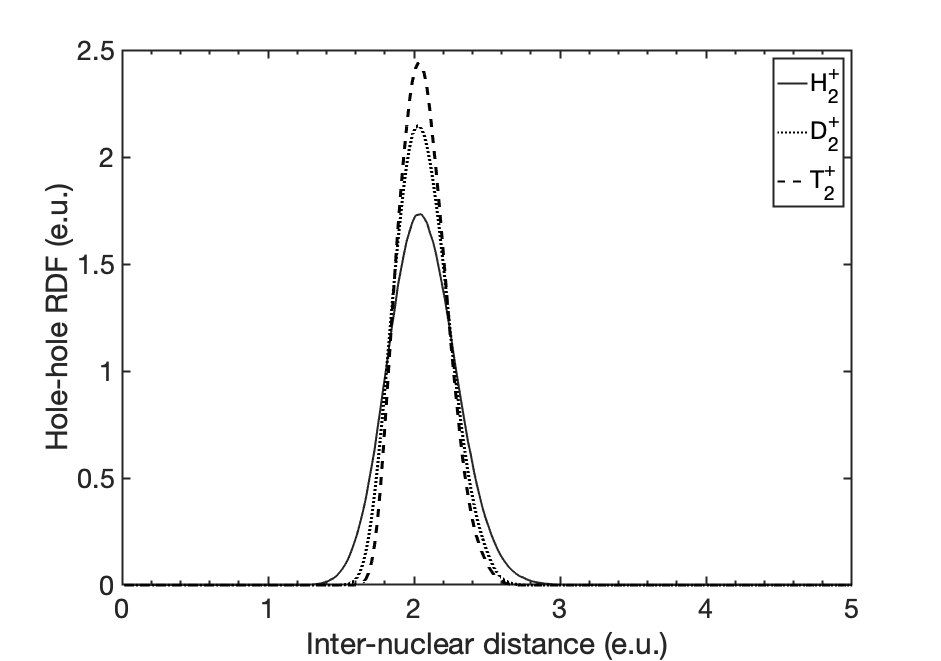}
\caption{Nucleus-nucleus RDF $4\pi r^2 g_\text{hh}(r)$ for three
  dihydrogen cations (H$_2^+$, D$_2^+$, and T$_2^+$) versus the
  nucleus-nucleus distance.  \label{fig:rdf_trions}}
\end{figure}

In another, more precise approach, we have calculated the mean
nucleus-nucleus distance of the molecular hydrogen cation from the
radial distribution function (RDF) $4\pi r^2 g_\text{hh}(r)$ obtained
using QMC calculations for the exact mass ratio, as shown in
Fig.\ \ref{fig:rdf_trions}.  The position of the peak in the hole-hole
RDF gives the most likely nucleus-nucleus distance for each
isotope. The mean nucleus-nucleus distance or bond length between two
nuclei can be evaluated as $\langle r_\text{hh} \rangle =
\int_0^\infty 4\pi r^2 g_\text{hh}(r)\, r \, dr$. DMC predicts a
slight increase in the bond length of the system when nuclear dynamics
are considered; see Table \ref{table:trion_rdf_spec}. Also, a slightly
larger equilibrium nucleus-nucleus distance for lighter isotopes is
obtained. These results are in good agreement with previous data when
nuclear dynamics are included in the calculations
\cite{Kylanpaa_2007}. In addition, the maximum of the RDF increases
with the nuclear mass as we approach the static-nucleus limit.

\begin{table*}[!htbp]
\centering
\caption{Adiabatic equilibrium nucleus-nucleus distance $R_\text{eq}$,
  nonadiabatic mean nucleus-nucleus and electron-nucleus distances
  $\langle r_\text{hh}\rangle$ and $\langle r_\text{eh}\rangle$, DMC
  zero-point energy (ZPE), difference $\Delta E_\text{Z}$ between the
  exact ZPE and the harmonic approximation $\omega_\text{e}/2$ to the
  ZPE, and spectroscopic constants of dihydrogen cations. Results are
  from the present work, except where citations are given.
\label{table:trion_rdf_spec}}
\begin{tabular}{lcccccccccc}
\hline \hline

Ion & $R_\text{eq}$ (a.u.) & $\langle r_\text{hh}\rangle$ (a.u.) &
$\langle r_\text{eh}\rangle$ (a.u.) & $\sigma_\text{hh}$ (a.u.) & ZPE
(eV) & $\Delta E_\text{Z}$ (meV) & $\omega_\text{e}$ (cm$^{-1}$) &
$\omega_\text{e}x_\text{e}$ (cm$^{-1}$) & $\alpha_\text{e}$
(cm$^{-1}$) & $B_\text{e}$ (cm$^{-1}$) \\

\hline

H$_2^+$ & \begin{tabular}{@{}c@{}} $1.9970(5)$
  \\ $1.9972(1)$\footnote{Data taken from
  Ref.\ \onlinecite{Schaad_1970}. \label{fn:ad_Schaad}}
  \\ $2.0$\footnote{Data taken from
  Ref.\ \onlinecite{Alexander_2005}. \label{fn:ad_Alexander}} \end{tabular}
& \begin{tabular}{@{}c@{}} $2.063(9)$ \\ $2.075(2)$\footnote{Data
    taken from Ref.\ \onlinecite{Kylanpaa_2007}.}
    \\ $2.06403(7)$\footnote{Data taken from
    Ref.\ \onlinecite{Bressanini_1997}.} \end{tabular}
& \begin{tabular}{@{}c@{}} $1.69(3)$ \\  \\ \\ \end{tabular} &
 \begin{tabular}{@{}c@{}}  $0.231(9)$ \\ \\ \\  \end{tabular} &  \begin{tabular}{@{}c@{}} $0.1500(5)$ \\ \\ \\ \end{tabular}  &  \begin{tabular}{@{}c@{}} $6.0(7)$ \\ \\ \\  \end{tabular} & \begin{tabular}{@{}c@{}}
 $2323(4)$ \\ $2315.3(6)$\footref{fn:ad_Alexander}
   \\ $2321$\footnote{Data taken from
   Ref.\ \onlinecite{Jefferts_1969}.\label{fn:ad_Jefferts}} \end{tabular}
 & \begin{tabular}{@{}c@{}} $68(4)$
     \\ $67.3(2)$\footref{fn:ad_Alexander}
     \\ $66.2$\footref{fn:ad_Jefferts} \end{tabular} &
\begin{tabular}{@{}c@{}} $1.67(3)$ \\ $1.600(3)$\footref{fn:ad_Alexander} \\
$1.68$\footref{fn:ad_Jefferts} \end{tabular}
& \begin{tabular}{@{}c@{}} $29.97$
    \\ $29.9626(2)$\footref{fn:ad_Alexander}
    \\ $30.2$\footref{fn:ad_Jefferts} \end{tabular} \\

\hline

D$_2^+$ & \begin{tabular}{@{}c@{}} $1.9970(5)$
  \\ $1.9972(1)$\footref{fn:ad_Schaad}
  \\ $2.0$\footref{fn:ad_Alexander} \end{tabular}
& \begin{tabular}{@{}c@{}} $2.059(9)$ \\ \\ \\  \end{tabular}
&  \begin{tabular}{@{}c@{}} $1.69(3)$ \\ \\ \\ \end{tabular} &
 \begin{tabular}{@{}c@{}} $0.180(9)$ \\ \\ \\ \end{tabular} &  \begin{tabular}{@{}c@{}} $0.106(1)$ \\ \\ \\ \end{tabular} & \begin{tabular}{@{}c@{}} $4(1)$ \\ \\ \\ \end{tabular} &
\begin{tabular}{@{}c@{}} $1643(3)$ \\ $1577.3$\footnote{Data taken from
 Ref.\ \onlinecite{Takezawa_1975}. \label{fn:Takezawa_1975}}
  \\  \\ \end{tabular} & \begin{tabular}{@{}c@{}}  $34(2)$
  \\ \\ \\ \end{tabular} & \begin{tabular}{@{}c@{}} $0.59(1)$
  \\ $0.560$\footref{fn:Takezawa_1975} \\ \\ \end{tabular}
& \begin{tabular}{@{}c@{}} $14.99$
    \\ $15.061$\footref{fn:Takezawa_1975} \\ \\ \end{tabular} \\

\hline

T$_2^+$ & \begin{tabular}{@{}c@{}} $1.9970(5)$
  \\ $1.9972(1)$\footref{fn:ad_Schaad} \end{tabular}
&\begin{tabular}{@{}c@{}} $2.061(9)$ \\ \\ \end{tabular}
& \begin{tabular}{@{}c@{}}$1.69(2)$ \\ \\ \end{tabular} &
\begin{tabular}{@{}c@{}} $0.159(9)$ \\ \\ \end{tabular} & \begin{tabular}{@{}c@{}} $0.087(1)$  \\ \\ \end{tabular} &  \begin{tabular}{@{}c@{}} $4(1)$ \\ \\ \end{tabular} & \begin{tabular}{@{}c@{}} $1343(2)$ \\ \\ \end{tabular} &  \begin{tabular}{@{}c@{}} $23(1)$ \\ \\ \end{tabular} &  \begin{tabular}{@{}c@{}} $0.322(6)$\\ \\ \end{tabular}  &
\begin{tabular}{@{}c@{}} $10.01$ \\ \\ \end{tabular} \\

\hline \hline
\end{tabular}
\end{table*}

\begin{figure}[!htbp]
\centering
\includegraphics[clip,width=1\linewidth]{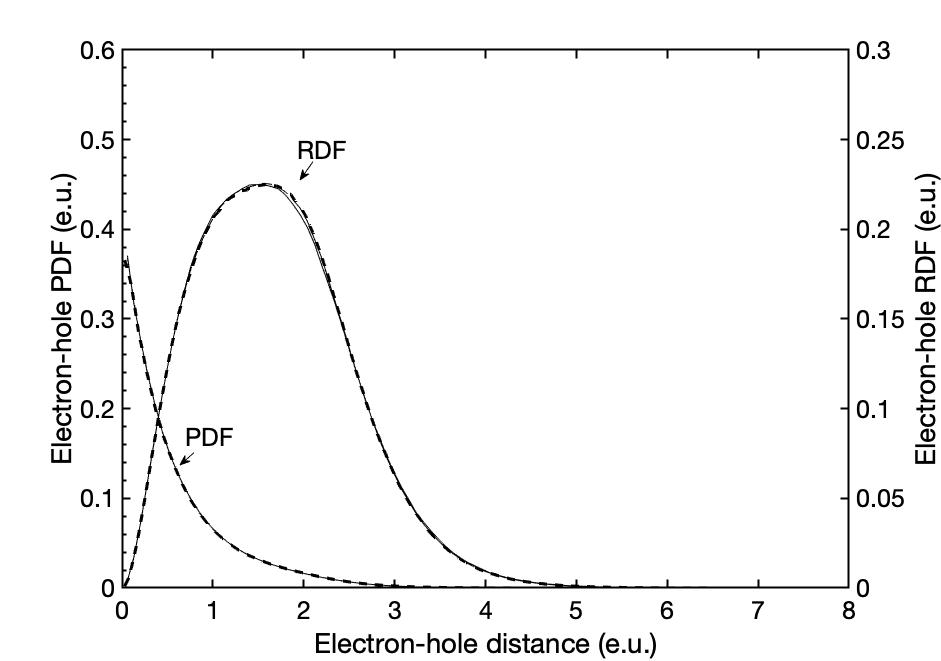}
\caption{Electron-nucleus PDF $g_\text{eh}(r)$ and the corresponding
  RDF $4\pi r^2 g_\text{eh}(r)$ in three dihydrogen cations (H$_2^+$,
  D$_2^+$, and T$_2^+$) versus the electron-nucleus distance. The
  curves for the different ions are almost on top of each
  other. \label{fig:pdf_trions}}
\end{figure}

The nucleus-nucleus spatial width of the lighter isotope H$_2^+$ is
slightly larger than the two others and shows a slight asymmetry,
demonstrating that anharmonicity effects are more important in H$_2^+$
than D$_2^+$ or T$_2^+$. The width of the RDF is quantified by the
standard deviation $\sigma_\text{hh}=\sqrt{\left\langle r_\text{hh}^2
  \right\rangle -\left\langle r_\text{hh}\right\rangle ^2}$.  The
standard deviation for H$_2^+$ is $0.231(9)$ a.u.\ and it decreases
for the more massive isotopes D$_2^+$ and T$_2^+$, as seen in Table
\ref{table:trion_rdf_spec}. A previous path integral Monte Carlo study
indicated a broadening of $0.539(1)$ a.u.\ and $0.454(1)$ a.u.\ at
half maximum of RDF digram for H$_2^+$ and D$_2^+$, respectively
\cite{Kylanpaa_2007}. On the other hand, as seen in
Fig.\ \ref{fig:pdf_trions}, isotope mass does not strongly influence
the electron-nucleus coupling, and we obtained three very similar
$g_\text{eh}(r)$ curves, as implied by the BO approximation. Figure
\ref{fig:pdf_trions} also shows that $g_\text{eh}(r)$ falls off
approximately exponentially with distance, with the associated
lengthscale being $0.58(1)$ e.u.

\subsubsection{Spectroscopic constants}

For a given electronic state, the spectrum of the atomic system is
determined by the corresponding vibrational and rotational levels. We
applied the BO potential energy curve described by
Eq.\ (\ref{eq:BO_fit}) to evaluate the contribution of rovibrational
motion to the total energy by evaluating the spectroscopic constants
of H$_2^+$ isotopes from a Dunham polynomial \cite{Dunham_1932}:
\begin{eqnarray}
E_{nJ} & = & E_{00} + \omega_\text{e}\left(n + \frac{1}{2} \right)
-\omega_\text{e}x_\text{e} {\left(n + \frac{1}{2} \right)}^2 \nonumber
\\ & & {} + B_\text{e} J(J + 1) - \alpha_\text{e} J(J + 1) \left(n +
\frac{1}{2} \right) + \ldots, \nonumber \\ \end{eqnarray} where
$E_{00}=U_\text{hh}(R_\text{eq})$ is the minimum of the BO potential,
$\omega_\text{e}$ is the harmonic vibration frequency about the
minimum of the BO potential, and $n=0, 1, 2, \ldots$ is the
vibrational quantum number. $\omega_\text{e} x_\text{e}$ describes the
effects of anharmonicity in the BO potential. $B_\text{e}J(J+1)$ is
the angular kinetic energy and $J= 0,1,2,\ldots$ is the rotational
quantum number. $\alpha_\text{e}$ describes the strength of
rovibrational coupling. In an adiabatic approach, we have used the
energy function introduced by Eq.\ (\ref{eq:BO_fit}) at the
equilibrium nucleus-nucleus distance within the BO approximation to
calculate the spectroscopic constants of the H$_2^+$ isotopes as
\cite{Dunham_1932}
\begin{eqnarray} B_\text{e} & = & \frac{1}{2\mu_\text{hh} r^2} \label{eq:B_e}
\\ \omega_\text{e} & = & {\left( \frac{1}{\mu_\text{hh}}
  \frac{d^2U_\text{hh}}{dr^2} \right)}^{1/2} \\  \omega_\text{e}
x_\text{e} & = & \frac{1}{48\mu_\text{hh}} \left[ 5 {\left(
    \frac{\frac{d^3U_\text{hh}}{dr^3}}{\frac{d^2U_\text{hh}}{dr^2}}
    \right)}^2 - 3
  \frac{\frac{d^4U_\text{hh}}{dr^4}}{\frac{d^2U_\text{hh}}{dr^2}}
  \right] \\ \alpha_\text{e} & = &
-\frac{6B_\text{e}^2}{\omega_\text{e}} \left( \frac{R}{3}
\frac{d^3U_\text{hh}/dr^3}{d^2U_\text{hh}/dr^2} + 1
\right). \label{eq:alpha_e} \end{eqnarray} Here, $r$ is the
nucleus-nucleus distance and $\mu_\text{hh}=m_\text{h}/2$ is the
reduced mass of the two nuclei. Our results are presented in Table
\ref{table:trion_rdf_spec} and compared with the available data in the
literature.

The spectroscopic parameters given in
Eqs.\ (\ref{eq:B_e})--(\ref{eq:alpha_e}) can be evaluated either at
the equilibrium separation $r=R_\text{eq}$ or by taking their
expectation values with respect to the nucleus-nucleus PDF\@.  The
results did not change significantly when the spectroscopic parameters
were calculated by taking their expectation value with respect to the
PDF instead of calculating the parameters at the adiabatic equilibrium
distance $R_\text{eq}$, and hence our reported results just used
$r=R_\text{eq}$.  In Table \ref{table:trion_energies} we report the
exact vibrational ZPE of each dihydrogen cation as the difference
between the ground-state energy and the minimum total energy of a
heavy-hole positive trion X$^+$ ($\sigma=0$). Comparing the ZPEs of
these three isotopes with the harmonic part of their vibrational
energy $\omega_\text{e}/2$ shows that anharmonicity makes a larger
contribution to the ZPE of H$_2^+$ than in the other two isotopes, as
can be seen in Table \ref{table:trion_rdf_spec}. As shown in Table
\ref{table:trion_rdf_spec}, the ZPE falls off as the nuclear mass
increases. Zero-point fluctuations are responsible for the broadening
of the nucleus-nucleus RDF in Fig.\ \ref{fig:rdf_trions} at
temperature $T=0$.

Electron-positron contact pair densities (which determine annihilation
rates) are given in Table \ref{table:elec_pos_pdf}.

\begin{table}[!htbp]
\caption{VMC-DMC extrapolated estimates of the opposite-spin
  electron-positron contact pair density $g_\text{eh}({\bf 0})/2$ in
  some positronic ions and molecules. The analytic result for Ps is
  shown for comparison.  The error bars quantify the uncertainty due
  to VMC and DMC simulation, but not errors due to the form and
  optimization of the trial wave function. \label{table:elec_pos_pdf}}
\begin{tabular}{lc}
\hline \hline

Complex & $g_\text{eh}({\bf 0})/2$ (a.u.) \\

\hline

Ps & $1/(16\pi)$ \\

Ps$^-$ & $0.020709(9)$ \\

Ps$_2$ & $0.04427(1)$ \\

PsH & $0.02465(2)$ \\

PsD & $0.02455(1)$ \\

PsT & $0.02457(1)$ \\

\hline \hline
\end{tabular}
\end{table}

\subsubsection{Accuracy of BO and harmonic approximations}

\begin{figure}[!htbp] 
  \centering
  \includegraphics[clip,width=1\linewidth]{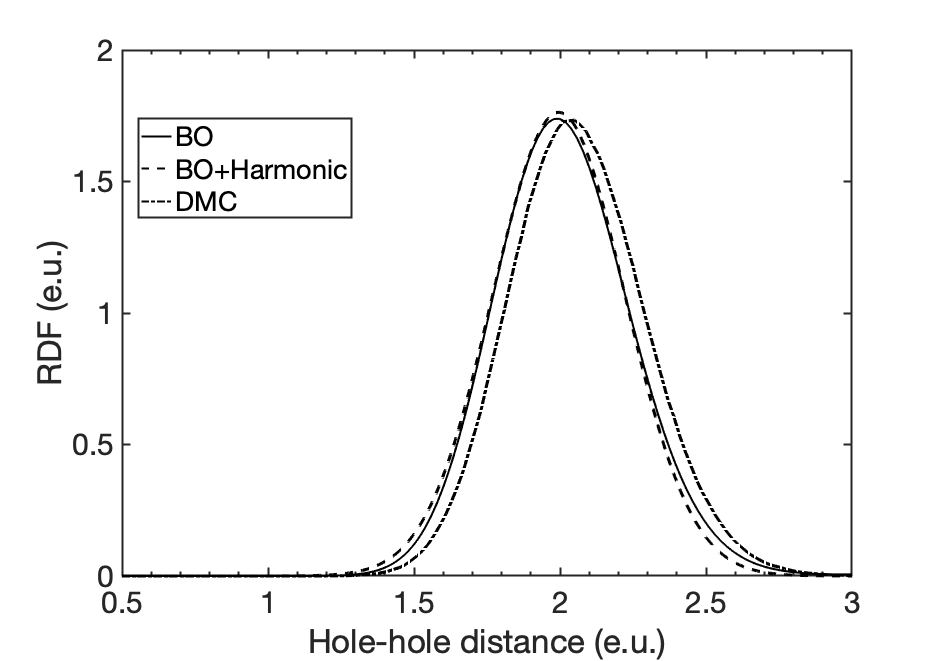}
  \caption{Proton-proton RDF in a hydrogen molecular cation obtained
    using the three different approaches listed in the caption of
    Table \ref{tab:shooting_energy}. \label{fig:RDF_shooting_trion}}
 \end{figure}

Equation (\ref{eq:BO_fit}) gives the BO potential energy surface,
which only depends on the hole-hole distance $r_\text{hh}$.
Transforming to the hole-hole center of mass and difference
coordinates, as done for electron-hole relative motion in an exciton
in Eq.\ (\ref{eq:X_SE}), the nuclear part of the Schr\"{o}dinger
equation within the BO approximation is
  \begin{equation} \left[-\frac{1}{2\mu_\text{hh}}\nabla_\text{hh}^2
+U_\text{hh}(r_\text{hh}) \right] \psi_\text{hh}({\bf
      r}_\text{hh})=E_\text{BO} \psi_\text{hh}({\bf
      r}_\text{hh}), \label{eq:schrodinger_holes} \end{equation} where
  $\psi_\text{hh}({\bf r}_\text{hh})$ is the hole-hole wave
  function. $E_\text{BO}$ represents the total energy of the system
  within the (fully anharmonic) BO approximation. In a spherically
  symmetric system, the wave function only depends on the hole-hole
  separation $r_\text{hh}$. Consequently,
  Eq.\ (\ref{eq:schrodinger_holes}) reduces to a one-dimensional
  Schr\"{o}dinger equation
  \begin{eqnarray} & & -\frac{1}{2\mu_\text{hh}} \left( \frac{d^2}{dr_\text{hh}^2}+\frac{2}{r}\dfrac{d}{dr_\text{hh}} \right) \psi_\text{hh}(r_\text{hh})+U_\text{hh}(r_\text{hh})
  \psi_\text{hh}(r_\text{hh}) \nonumber \\ & & {} = E_\text{BO}
  \psi_\text{hh}(r_\text{hh}). \label{eq:1D_schrodinger_holes} \end{eqnarray}
  We have employed the shooting method with the initial condition
  $\psi_\text{hh}(0)=0$ and
  $\frac{d\psi_\text{hh}(r_\text{hh})}{dr_\text{hh}}=0$ at large
  separation to solve Eq.\ (\ref{eq:1D_schrodinger_holes}) numerically
  for the H$_2^+$ cation. In Table \ref{tab:shooting_energy} we
  compare (i) the exact nonrelativistic energy obtained using DMC for
  all the constituent particles, (ii) the energy within the fully
  anharmonic BO approximation obtained using the shooting method, and
  (iii) the harmonic approximation within the BO framework, in which
  the total energy is
  \begin{equation} E_\text{BO+Harm.} = U_\text{hh}(R_\text{eq})
+\frac{\omega_\text{e}}{2}.
\end{equation}
 From the data in Table \ref{tab:shooting_energy}, the BO energies are
 about 0.0003--0.0005 a.u.\ lower than the DMC-calculated exact
 energies. In Fig.\ \ref{fig:RDF_shooting_trion} the probability
 density $|\psi_\text{hh}(r_\text{hh})|^2$ within the BO approximation
 is compared with both the DMC hole-hole RDF and the ground state
 probability density in the harmonic approximation, ${(\mu_\text{hh}
   \omega_\text{e}/ \pi)}^{1/2} e^{-\mu_\text{hh} \omega_\text{e}
   {(r_\text{hh}-R_\text{eq})}^2}$, where $R_\text{eq}$ and
 $\omega_\text{e}$ are taken from Table
 \ref{table:trion_rdf_spec}. The BO fully anharmonic hole-hole RDF,
 shown in Fig.\ \ref{fig:RDF_shooting_trion}, is in slightly better
 agreement with the DMC hole-hole RDF than is the BO harmonic
 approximation, especially for the tails of the RDF\@.  On the
 other hand, surprisingly, the harmonic approximation gives a more
 accurate ground-state total energy than the fully anharmonic BO
 approximation in comparison with the DMC simulation of all the
 particles.  The likely reason for this pathological behavior of the
 BO approximation is that, at the level of accuracy at which we are
 working, the ambiguity in the mass of the heavy particles is
 significant: should some fraction of the electron mass be included in
 $\mu_\text{hh}$?
 
 \begin{table}[!htbp]
  \centering
  \caption{Total energy of the molecular hydrogen cation H$_2^+$ and
    dihydrogen H$_2$ from three different approaches: (i) the exact
    nonrelativistic solution $E$ obtained using DMC, (ii) the BO
    approximation $E_\text{BO}$, (iii) the BO and harmonic
    approximations $E_\text{BO+Harm.}$ \label{tab:shooting_energy}}
  \begin{tabular}{lccc}
      \hline \hline

   Complex & $E$ (a.u.) & $E_\text{BO}$ (a.u.) &  $E_\text{BO+Harm.}$
   (a.u.) \\

    \hline 

   H$_2^+$ & $-0.5971384(7)$ & $-0.59738903$ & $-0.597340(7)$ \\

   H$_2$ & $-1.164015(8)$ & $-1.1645088$ & $-1.16441(2)$ \\
   
\hline \hline
\end{tabular} 
 \end{table}
 
\subsection{Biexcitons}

\subsubsection{Binding energies}

The binding energy of a biexciton is the energy required to decompose
it into its two constituent excitons:
\begin{equation} E_{\text{X}_2}^\text{B}=2E_\text{X}-E_{\text{X}_2},
\end{equation}
where $E_{\text{X}_2}$ is the ground state total energy of the
biexciton.

As we have done for a trion, in the limit of two heavy holes, we can
employ the BO approximation and separate the vibrational contribution
of heavy holes to the total energy from the electronic
contribution. The derivation of Eq.\ (\ref{eq:harmonic_approx}) is
exactly the same for a biexciton as for a positive trion, again
leading to the conclusion that the binding energy must go as
$\sqrt{x}$ at small $x$. However, unlike a trion, the binding energy
must be unchanged under the exchange of electrons and holes (i.e.,
$m_\text{e} \leftrightarrow m_\text{h}$ or, equivalently, $\sigma
\leftrightarrow 1/\sigma$ or $x \leftrightarrow 1-x$). This suggests
that a suitable fitting function for the binding energy of a biexciton
should be a symmetric polynomial in $\sqrt{x}$ and $\sqrt{1-x}$. We
found the DMC binding energy data to be well fitted by
\begin{equation} E_{\text{X}_2}^\text{B}=\sum_{i=0}^4 c_i {[x(1-x)]}^{i/2}. \label{eq:biex_fit}
\end{equation}
The fitting parameters $\{c_i\}$ are presented in Table
\ref{table:biex_fit_params} and the fitted curve is plotted along with
the raw data in Fig.\ \ref{fig:be_biex_fit}. The DMC total energies
are listed in Appendix \ref{app:raw_data}. The SSE and RMSE are $2.249
\times 10^{-9}$ e.u.\ and $1.224\times 10^{-5}$ e.u., respectively.

\begin{table}[!htbp]
\centering
\caption{Fitted parameters in Eq.\ (\ref{eq:biex_fit}) for biexciton
  binding energies. \label{table:biex_fit_params}}
\begin{tabular}{lc}
\hline \hline

Parameter & Value (e.u.) \\

\hline

$c_0$ & $0.17438546591410939$ \\

$c_1$ & $-0.42831856512902888$ \\

$c_2$ & $0.38685906773221207$ \\

$c_3$ & $-0.26511242810730029$ \\

$c_4$ & $0.13132632976660738$ \\

\hline \hline
\end{tabular}
\end{table}

\begin{figure}[!htbp]
\centering
\includegraphics[clip,width=1\linewidth]{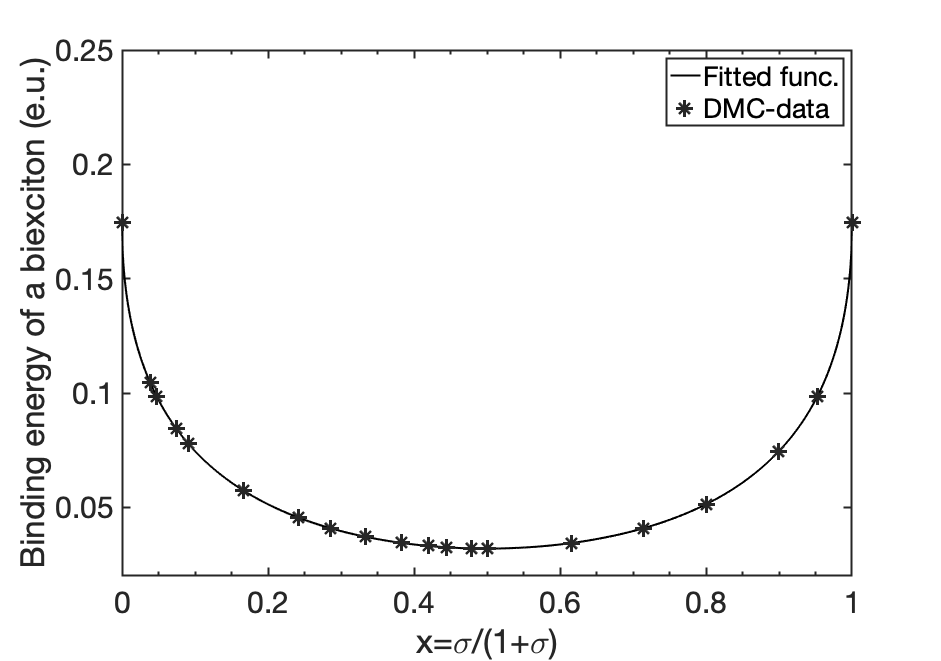}
\caption{DMC binding energies of biexcitons against rescaled mass
  ratio $x=\sigma/(1+\sigma)$. The solid line shows the fit of
  Eq.\ (\ref{eq:biex_fit}). 
\label{fig:be_biex_fit}}
\end{figure}

Using Eq.\ (\ref{eq:biex_fit}), we predict the biexciton binding
energy in various semiconductors in Table \ref{table:trion_BEs}. In
all cases, the biexciton binding energies are larger than the trion
binding energies, but are of the same order of magnitude.

\subsubsection{Mass effects in molecular hydrogen isotopes}

An H$_2$ molecule consists of two protons and two electrons and hence
resembles a biexciton with heavy holes. Likewise, a dimuonium molecule
Mu$_2$ is formed by the electrostatic interaction between two
electrons and two antimuons. The binding energy of such a molecule is
defined as the energy required to dissociate the molecule into two
isolated atoms. The DMC ground state total energies and binding
energies of dimuonium and the molecular hydrogen isotopes H$_2$,
D$_2$, and T$_2$ are given in Table \ref{table:biex_energies}.  The
DMC-calculated total energies are in excellent agreement with previous
quantum electrodynamics predictions \cite{Puchalski_2019}, confirming
that the molecules are well described by nonrelativistic quantum
mechanics. Indeed, the DMC binding energies retrieve the experimental
outcomes and show that the binding energy increases with nuclear
mass. As can be seen in Table \ref{table:biex_energies},
Eq.\ (\ref{eq:biex_fit}) predicts the binding energies of the three
molecular hydrogen isotopes in good agreement with both the raw DMC
data and experimental results.

\begin{table*}[!htbp]
\centering
\caption{As Table \ref{table:trion_energies}, but for neutral
  dihydrogen-like molecules. Raw data in
  Ref.\ \onlinecite{Puchalski_2019} are given in a.u.; for easy
  comparison with DMC results, we converted them to e.u.\ and rounded
  them to a smaller number of digits.
\label{table:biex_energies}}
\begin{tabular}{lcccccc}
\hline \hline

\multirow{2}{*}{Molecule} & \multicolumn{3}{c}{Total energy (e.u)} &
\multicolumn{3}{c}{Binding energy (eV)} \\

& DMC & Eq.\ (\ref{eq:biex_fit}) & Prev.\ works & DMC &
Eq.\ (\ref{eq:biex_fit}) & Exp. \\

\hline

Ps$_2$ (e$^-$e$^-$e$^+$e$^+$) & $-1.032009(3)$ & $-1.032010(3)$ &
$-1.03196$\footnote{Data taken from
Refs.\ \onlinecite{kozlowski1993nonadiabatic}} & $0.43550(4)$ &
$0.43552(4)$ \\

Mu$_2$ (e$^-$e$^-\mu^+\mu^+$) & $-1.1465(1)$ & $-1.14651(1)$ & &
$3.967(3)$ & $3.9676(3)$ & \\

H$_2$ (e$^-$e$^-$p$^+$p$^+$) & $-1.164649(8)$ & $-1.16460(1)$ &
$-1.164659$\footnote{Data taken from
Refs.\ \onlinecite{Puchalski_2019} and
\onlinecite{puchalski2019nonadiabatic}
(QED)\@. \label{fn:Etot_Puchalski}} & $4.4779(2)$ & $4.4766(3)$ &
$4.478$\footnote{Data taken from
Ref.\ \onlinecite{Liu_2010}. \label{fn:Etot_Liu}} \\

D$_2$ (e$^-$e$^-$d$^+$d$^+$) & $-1.167482(2)$ & $-1.16742(1)$ &
$-1.167487$\footref{fn:Etot_Puchalski} & $4.55618(5)$ & $4.5545(3)$ &
$4.556$\footref{fn:Etot_Liu} \\

T$_2$ (e$^-$e$^-$t$^+$t$^+$) & $-1.168749(3)$ & $-1.16868(1)$ &
$-1.168748$\footref{fn:Etot_Puchalski} & $4.59106(8)$ & $4.5892(3)$ &
$4.586$\footref{fn:Etot_Puchalski} \\

X$_2$ (e$^-$e$^-$h$^+$h$^+$), with $\sigma=0$ & $-1.17443(3)$ &
$-1.17439(1)$ & &  $4.7465(8)$ & $4.7454(3)$ & \\

\hline \hline
\end{tabular}
\end{table*}

\subsubsection{BO potential energy curve}

\begin{figure}[!htbp]
\centering
\includegraphics[clip,width=1\linewidth]{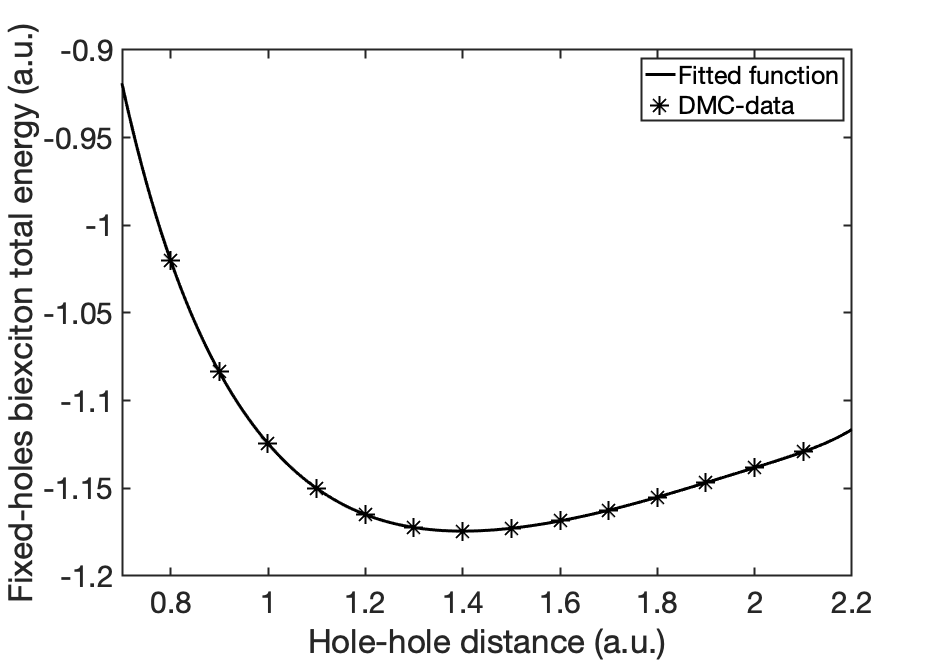}
\caption{DMC BO potential energy, i.e., DMC total energy of a
  biexciton in the infinite mass limit of two heavy holes and two
  light electrons ($\sigma=0$) against the hole-hole distance. The
  solid curve shows a fit of Eq.\ (\ref{eq:biex_BO}) to the DMC data.
}
\label{fig:bo_biexciton}
\end{figure}

As with molecular cations, the BO potential energy curve provides
important information about the nuclear motion around the equilibrium
separation. We performed a series of DMC simulations of two electrons
in the presence of two fixed holes at different hole-hole distances,
between 0.8 and 1.8 a.u. The results are shown in Appendix
\ref{app:raw_data} and compared with the available previous data. We
found that a polynomial of degree eight,
\begin{equation} U_\text{hh}(r)=\sum_{i=0}^8 p_{i}r^i, \label{eq:biex_BO}
\end{equation}
 fitted our data excellently, with the coefficients $\{p_i\}$ reported
 in Appendix \ref{app:dmc_bo_data}. The SSE and RMSE are $3.02 \times
 10^{-9}$ and $3.173 \times 10^{-5}$ a.u., respectively. Figure
 \ref{fig:bo_biexciton} shows how well the curve given by
 Eq.\ (\ref{eq:biex_BO}) passes through the DMC data.

\subsubsection{Spectroscopic constants}

We have calculated the spectroscopic constants of molecular hydrogen
from the potential energy curve given by Eq.\ (\ref{eq:biex_BO}),
using Eqs.\ (\ref{eq:B_e})--(\ref{eq:alpha_e}) evaluated at the
equilibrium separation. The results are shown in Table
\ref{table:biex_spect}. For comparison, the spectroscopic constants
obtained from a Morse model are given in Appendix \ref{app:bo_fits}.
The adiabatic equilibrium nucleus-nucleus distance is
$R_\text{eq}=1.4$ a.u., which agrees with a previous adiabatic VMC
prediction \cite{Alexander_2004}. Furthermore, we obtained the
vibrational ZPE of these molecules by comparing adiabatic and
nonadiabatic energies. Our spectroscopic constants are in good
agreement with the available experiments, as seen in Table
\ref{table:biex_spect}.
 
\begin{table*}[!htbp]
\centering
\caption{As Table \ref{table:trion_rdf_spec}, but for neutral
  dihydrogen molecules. \label{table:biex_spect}}
\begin{tabular}{lcccccccccc}
\hline \hline

Mol. & $R_\text{eq}$ (a.u.) & $\langle r_\text{hh}\rangle$ (a.u.) &
$\langle r_\text{eh}\rangle$ (a.u.)  & $\sigma_\text{hh}$ (a.u.)  &
ZPE (eV) & $\Delta E_\text{Z}$ (meV) & $\omega_\text{e}$ (cm$^{-1}$) &
$\omega_\text{e}x_\text{e}$ (cm$^{-1}$) & $\alpha_\text{e}$
(cm$^{-1}$) & $B_\text{e}$ (cm$^{-1}$) \\

\hline

H$_2$ & \begin{tabular}{@{}c@{}} $1.400(5)$ \\ $1.4$\footnote{Data
  taken from Ref.\ \onlinecite{Alexander_2004}.}
  \\ $1.401$\footnote{Data taken from
  Ref.\ \onlinecite{Puchalski_2019}.} \end{tabular} &
\begin{tabular}{@{}c@{}} $1.44(2)$ \\ $1.4193$\footnote{Data taken from
Ref.\ \onlinecite{Stoicheff_1957} (experimental, no error bars
given). \label{fn:Etot_Stoicheff}} \\ \\ \end{tabular}
&  \begin{tabular}{@{}c@{}} $1.57(6)$ \\ \\ \\ \end{tabular}  &
\begin{tabular}{@{}c@{}} $0.16(2)$ \\ \\ \\ \end{tabular} & \begin{tabular}{@{}c@{}} $0.2834(8)$
  \\ $0.270$\footnote{Data taken from Ref.\ \onlinecite{Irikura_2007}
  (experimental). \label{fn:dist_Irikura}} \\ \\  \end{tabular}
&  \begin{tabular}{@{}c@{}} $10(1)$ \\ \\  \\ \end{tabular} &
\begin{tabular}{@{}c@{}} $4400(10)$ \\ $4401.213$\footref{fn:dist_Irikura} \\ \\ \end{tabular} & \begin{tabular}{@{}c@{}} $110(30)$ \\
$121.336$\footref{fn:dist_Irikura} \\  \\ \end{tabular}
& \begin{tabular}{@{}c@{}} $3.1(2)$
    \\ $3.0622$\footref{fn:dist_Irikura} \\ \\ \end{tabular}
& \begin{tabular}{@{}c@{}} $60.984$
    \\ $60.853$\footref{fn:dist_Irikura} \\ \\ \end{tabular} \\

\hline

D$_2$ &  \begin{tabular}{@{}c@{}} $1.400(5)$ \\ \\ \end{tabular}
& \begin{tabular}{@{}c@{}} $1.43(2)$
    \\ $1.4139$\footref{fn:Etot_Stoicheff} \end{tabular}
&  \begin{tabular}{@{}c@{}} $1.57(4)$  \\ \\ \end{tabular}  &
\begin{tabular}{@{}c@{}} $0.14(2)$  \\ \\ \end{tabular}& \begin{tabular}{@{}c@{}} $0.1977(8)$
 \\ $0.1917$\footref{fn:Etot_Stoicheff} \end{tabular}
&  \begin{tabular}{@{}c@{}} $4(1)$ \\ \\ \end{tabular}
& \begin{tabular}{@{}c@{}} $3120(10)$
    \\ $3115.5$\footref{fn:dist_Irikura} \end{tabular}
& \begin{tabular}{@{}c@{}} $55(7)$
    \\ $61.82$\footref{fn:dist_Irikura} \end{tabular}
& \begin{tabular}{@{}c@{}} $1.03(7)$
    \\ $1.0786$\footref{fn:dist_Irikura} \end{tabular}
& \begin{tabular}{@{}c@{}} $30.50739(2)$
    \\ $30.4436$\footref{fn:dist_Irikura} \end{tabular} \\

\hline

T$_2$ & \begin{tabular}{@{}c@{}} $1.400(5)$ \\ \\  \end{tabular}
& \begin{tabular}{@{}c@{}} $1.43(2)$ \\ $1.41146(4)$\footnote{Data
    taken from Ref.\ \onlinecite{Edwards_1978}.} \end{tabular}
&  \begin{tabular}{@{}c@{}} $1.57(6) $ \\ \\  \end{tabular} &
 \begin{tabular}{@{}c@{}} $0.13(2)$\\ \\  \end{tabular} & \begin{tabular}{@{}c@{}} $0.1604(8) $
  \\ $0.1569$\footnote{Data taken from Ref.\ \onlinecite{Cashion_1966}
  (theoretical). \label{fn:dist_Cashion}} \end{tabular}
 & \begin{tabular}{@{}c@{}} $2(1)$ \\ \\  \end{tabular}
 & \begin{tabular}{@{}c@{}} $2547(8)$
     \\ $2546.4$\footref{fn:dist_Cashion} \end{tabular}
 & \begin{tabular}{@{}c@{}} $36(5)$
     \\ $41.23$\footref{fn:dist_Cashion} \end{tabular}
 & \begin{tabular}{@{}c@{}} $0.56(4)$ \\ $0.5887$\footnote{Data taken
     from Ref.\ \onlinecite{Kolos_1969}.} \end{tabular}
 & \begin{tabular}{@{}c@{}} $20.371$
     \\ $20.335$\footref{fn:dist_Cashion} \end{tabular} \\

\hline \hline
\end{tabular}
\end{table*}

In Appendix \ref{app:bo_fits} we compare our model, given by
Eq.\ (\ref{eq:biex_BO}), with the Morse potential. As in the case of
trions, although a Morse potential behaves much better than
Eq.\ (\ref{eq:biex_BO}) at large hole-hole separations, it does not
match the DMC data so well in the vicinity of the equilibrium point.

\begin{figure}[!htbp]
\centering
\includegraphics[clip,width=1\linewidth]{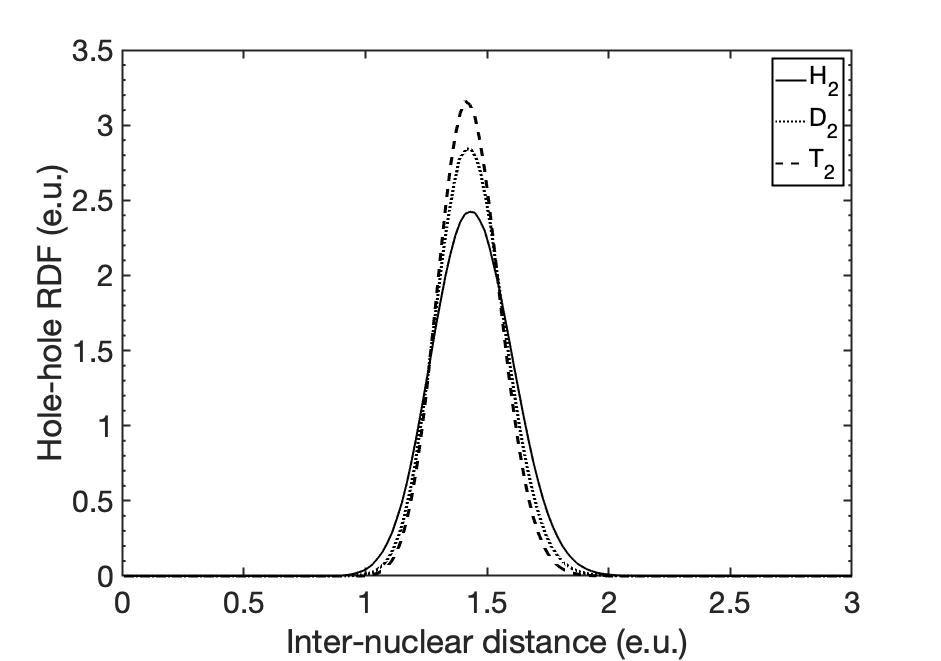}
\caption{Nucleus-nucleus RDF $4\pi r^2 g_\text{hh}(r)$ for three
  dihydrogen molecules (H$_2$, D$_2$, and T$_2$) versus the
  nucleus-nucleus distance. }
\label{fig:biex_hh_rdf}
\end{figure} 

\begin{figure}[!htbp]
\centering
\includegraphics[clip,width=1\linewidth]{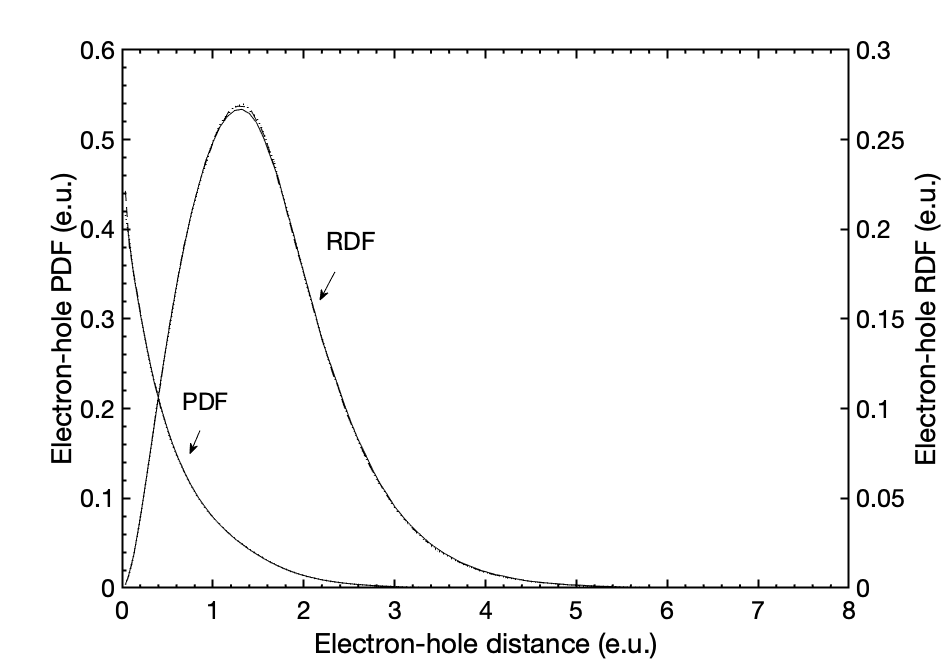}
\caption{Electron-nucleus PDF $g_\text{eh}(r)$ and RDF $4\pi r^2
  g_\text{eh}(r)$ in three dihydrogen molecules (H$_2$, D$_2$, and
  T$_2$) versus the electron-nucleus distance.}
\label{fig:rdf_biex_eh}
\end{figure}

Using the RDF results obtained from nonadiabatic QMC simulations of
two electrons and two nuclei, we explore the effects of nuclear motion
on the bond lengths of the molecular hydrogen isotopes. Our results
show slight increases in bond length when nuclear dynamics are
considered, bringing our results into agreement with the available
experimental data, as shown in Table \ref{table:biex_spect}.

Figure \ref{fig:biex_hh_rdf} shows the nucleus-nucleus RDF for the
three molecular hydrogen isotopes. As is the case for H$_2^+$, the
distribution of nucleus-nucleus distances in H$_2$ has a larger
spatial broadening than in the more massive isotopes, as quantified by
the standard deviation of the nucleus-nucleus distance reported in
Table \ref{table:biex_spect}. Accordingly, the ZPE increases with
electron-hole mass ratio $\sigma$, as seen in Table
\ref{table:biex_spect}. The effects of anharmonicity are visible in
Fig.\ \ref{fig:biex_hh_rdf} as an asymmetry in the RDF\@.

On the other hand, isotope mass does not influence the
electron-nucleus coupling: the three molecules show the same
electron-nucleus PDF curves in Fig.\ \ref{fig:rdf_biex_eh}. Indeed,
for all three molecules, the mean electron-nucleus distance $\langle
r_\text{eh} \rangle$ is only slightly less than the mean
electron-nucleus distance in the molecular cations.

\begin{figure}[!htbp] 
\centering
\includegraphics[clip,width=1\linewidth]{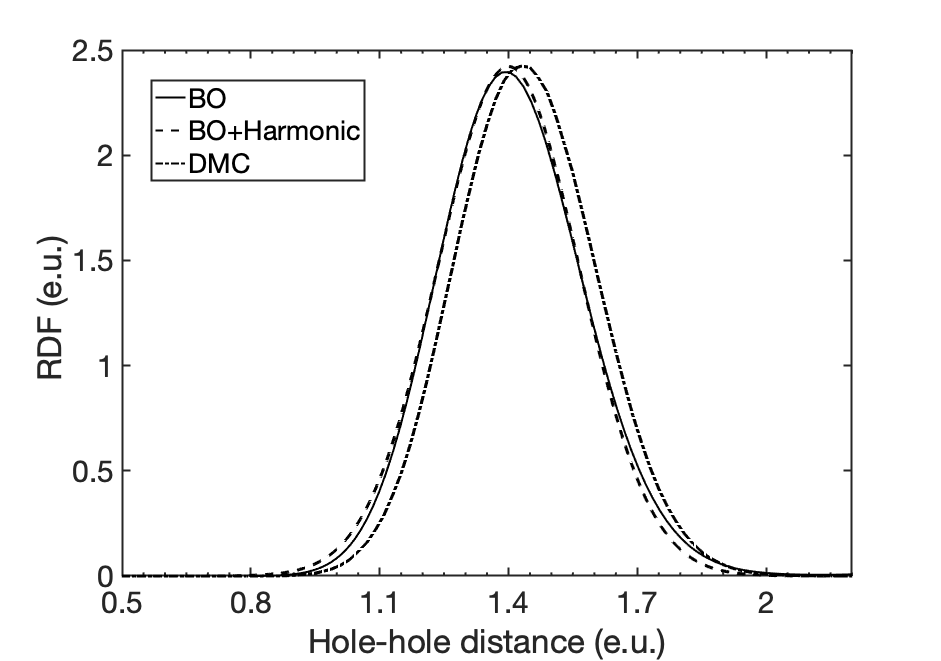}
\caption{Proton-proton RDF in a hydrogen molecule H$_2$ from three
  different approaches: BO approximation, BO+harmonic approximation,
  and the DMC solution to the four-particle
  problem. \label{fig:RDF_shooting_biexciton}}
\end{figure}

The exact nonrelativistic energy of H$_2$, the energy within the BO
approximation, and the energy within the harmonic approximation within
the BO framework are shown in Table \ref{tab:shooting_energy} and the
corresponding RDFs are shown in
Fig.\ \ref{fig:RDF_shooting_biexciton}.  The mean separation obtained
within exact DMC calculations for all four particles is greater than
the equilibrium separation of the BO potential.  As in the dihydrogen
cation, the harmonic approximation appears to perform better than the
fully anharmonic BO approximation. 

\subsection{Other Coulomb complexes: mixed hydrogenic molecules and cations;
helium hydride; and positronic and muonic complexes}

The DMC-calculated nonrelativistic ground-state total energies of a
number of small Coulomb complexes are shown in Table
\ref{table:misc_complexes}.  We take the masses of a proton (p$^+$),
deuteron (d$^+$), triton (t$^+$), helion (h$^{2+}$), alpha particle
($\alpha^{2+}$), and muon ($\mu^\pm$) to be
$m_{\text{p}^+}=1836.152673440001$ a.u.,
$m_{\text{d}^+}=3670.482967853717$ a.u.,
$m_{\text{t}^+}=5496.921535729647$ a.u.,
$m_{\text{h}^{2+}}=5495.885280115730$ a.u.,
$m_{\alpha^{2+}}=7294.299425443957$ a.u., and
$m_{\mu^\pm}=206.7682830910218$ a.u., respectively
\cite{CODATA_2018}. With the exception of H$_3^+$, the ground-state
spatial wave function is nodeless in each case and hence DMC provides
numerically exact solutions to the nonrelativistic Schr\"{o}dinger
equation.  For H$_3^+$ we treat the three protons as distinguishable
particles.  Many of these compounds play important roles in
interstellar chemistry.

\begin{table*}[!htbp]
\caption{DMC nonrelativistic ground-state total energies of various
  Coulomb complexes.
\label{table:misc_complexes}}
\begin{tabular}{lcc}
\hline \hline

\multirow{2}{*}{Complex} & \multicolumn{2}{c}{Total energy (a.u.)} \\

& DMC & Prev.\ works \\

\hline

$\mu^+$H (e$^-\mu^+$p$^+$) & $-0.58990(2)$ \\

$\mu^+$D (e$^-\mu^+$d$^+$) & $-0.59023(2)$ \\

$\mu^+$T (e$^-\mu^+$t$^+$) & $-0.59028(3)$ \\

\hline

HD (e$^-$e$^-$p$^+$d$^+$) & $-1.165472(7)$ &
$-1.16547192396366(5)$\footnote{Data taken from
Ref.\ \cite{Puchalski_2019} \label{fn:puchalski2019}} \\

HT (e$^-$e$^-$p$^+$t$^+$) & $-1.166027(9)$ &
$-1.16600203732867(6)$\footref{fn:puchalski2019} \\

DT (e$^-$e$^-$d$^+$t$^+$) & $-1.16781(1)$ &
$-1.16781967343673(5)$\footref{fn:puchalski2019} \\

\hline

HD$^+$ (e$^-$p$^+$d$^+$) & $-0.59790(4)$ \\

HT$^+$ (e$^-$p$^+$t$^+$) & $-0.59817(1)$ \\

DT$^+$ (e$^-$d$^+$t$^+$) & $-0.59915(2)$ \\

\hline

$^3$He$_2^{2+}$ (e$^-$e$^-$h$^{2+}$h$^{2+}$ & $-3.672410(7)$ \\

$^3$He\,$^4$He$^{2+}$ (e$^-$e$^-$h$^{2+}\alpha^{2+}$) & $-3.67302(1)$
\\

$^4$He$_2^{2+}$ (e$^-$e$^-\alpha^{2+}\alpha^{2+}$) & $-3.67364(1)$ \\

\hline

H$_3^+$ (e$^-$e$^-$p$^+$p$^+$p$^+$) & $-1.32344(1)$ \\

H$_2$D$^+$ (e$^-$e$^-$p$^+$p$^+$d$^+$) & $-1.325273(7)$ \\

H$_2$T$^+$ (e$^-$e$^-$p$^+$p$^+$t$^+$) & $-1.325985(9)$ \\

HD$_2^+$ (e$^-$e$^-$p$^+$d$^+$d$^+$) & $-1.327270(9)$ \\

HDT$^+$ (e$^-$e$^-$p$^+$d$^+$t$^+$) & $-1.328035(9)$ \\

HT$_2^+$ (e$^-$e$^-$p$^+$t$^+$t$^+$) & $-1.328830(9)$ \\

D$_3^+$ (e$^-$e$^-$d$^+$d$^+$d$^+$) & $-1.329399(9)$ \\

D$_2$T$^+$ (e$^-$e$^-$d$^+$d$^+$t$^+$) & $-1.330230(9)$ \\

DT$_2^+$ (e$^-$e$^-$d$^+$t$^+$t$^+$) & $-1.331115(9)$ \\

T$_3^+$ (e$^-$e$^-$t$^+$t$^+$t$^+$) & $-1.332040(9)$ \\

\hline

$^3$He (e$^-$e$^-$h$^{2+}$) & $-2.9031670(7)$ \\

$^4$He (e$^-$e$^-\alpha^{2+}$) & $-2.9033053(7)$ \\

\hline

$^3$HeH$^+$ (e$^-$e$^-$p$^+$h$^{2+}$) & $-2.970719(6)$ \\

$^4$HeH$^+$ (e$^-$e$^-$p$^+\alpha^{2+}$) & $-2.97107(1)$ \\

$^3$HeD$^+$ (e$^-$e$^-$d$^+$h$^{2+}$) & $-2.972272(9)$ \\

$^4$HeD$^+$ (e$^-$e$^-$d$^+\alpha^{2+}$) & $-2.972691(7)$ \\

$^3$HeT$^+$ (e$^-$e$^-$t$^+$h$^{2+}$) & $-2.972895(7)$ \\

$^4$HeT$^+$ (e$^-$e$^-$t$^+\alpha^{2+}$) & $-2.97335(1)$ \\

\hline

PsH (e$^-$e$^-$e$^+$p$^+$) & $-0.78890(2)$ \\

PsD (e$^-$e$^-$e$^+$d$^+$) & $-0.78904(3)$ \\

PsT (e$^-$e$^-$e$^+$t$^+$) & $-0.78909(3)$ \\

\hline \hline
\end{tabular}
\end{table*}

For the positronic compounds we report the contact pair density
$g_\text{eh}({\bf 0})/2$ between spin-up electrons and the positron in
Table \ref{table:elec_pos_pdf}.  It is clear that isotope effects in
the electron-positron annihilation rate in PsH, PsD, and PsT are
small.

\section{Conclusions}

We report high-precision, statistically exact DMC calculations of the
binding energies of three-dimensional excitonic complexes in terms of
the electron-hole mass ratio. In particular, we have focused on three-
and four-body complexes (trions and biexcitons) formed from
distinguishable electrons and holes with isotropic effective masses
and interacting via an isotropic $1/r$ Coulomb potential. Based on our
DMC data, we obtained interpolation formulas for the binding energies
of trions and biexcitons. These formulas can be applied to interpret
experimental photoabsorption and photoluminescence spectra in 3D
semiconductors. Furthermore, based on DMC calculations with small mass
ratios, we have calculated the nonrelativistic binding energies of
``real'' three-, four- and five-particle Coulomb complexes, including
hydrogen molecules and ions (with different isotopes), helium hydride
cations, and small positronic and muonic complexes. Using QMC PDFs, we
predict the nonadiabatic nucleus-nucleus distance, nuclear spatial
distribution, and spectroscopic constants for hydrogen molecules and
ions.  Where comparison is possible, our nonrelativistic results are
in good agreement with both experiments and previous theoretical
results obtained within quantum electrodynamics.

We find reasonable agreement between the exact nonrelativistic total
energy of H$_2$ and the total energy within the BO
approximation. Interestingly, we find that the total energy evaluated
within the BO approximation including all anharmonic effects is
slightly less accurate than the total energy evaluated using the
harmonic approximation to the BO potential.  Similar conclusions are
reached for H$_2^+$.  This shows that it cannot be guaranteed that the
inclusion of vibrational anharmonicity within the BO framework
improves the total energy of a molecule or crystal, especially when
light atoms such as hydrogen are present.

An important conclusion of our work is that, at least for the small
molecules that we have studied, QMC simulations in which the nuclei
are treated as quantum particles on an equal footing with the
electrons are no more difficult and only proportionately more
expensive than calculations with fixed nuclei, provided that an
appropriate vibrational Jastrow factor is used. This holds out the
prospect that QMC calculations for larger systems with a full quantum
treatment of nuclei could routinely be performed using Jastrow factors
that are quadratic functions of the phonon normal coordinates.

\begin{acknowledgments}
We acknowledge useful conversations with A.\ Hills, E.\ Cully, and
M.\ Szyniszewski.
\end{acknowledgments}

\bibliography{complexes} 

\appendix

\section{Extrapolation of DMC results to zero time step} \label{app:time_step}

\begin{figure}[!htbp]
 \includegraphics[clip,width=1\linewidth]{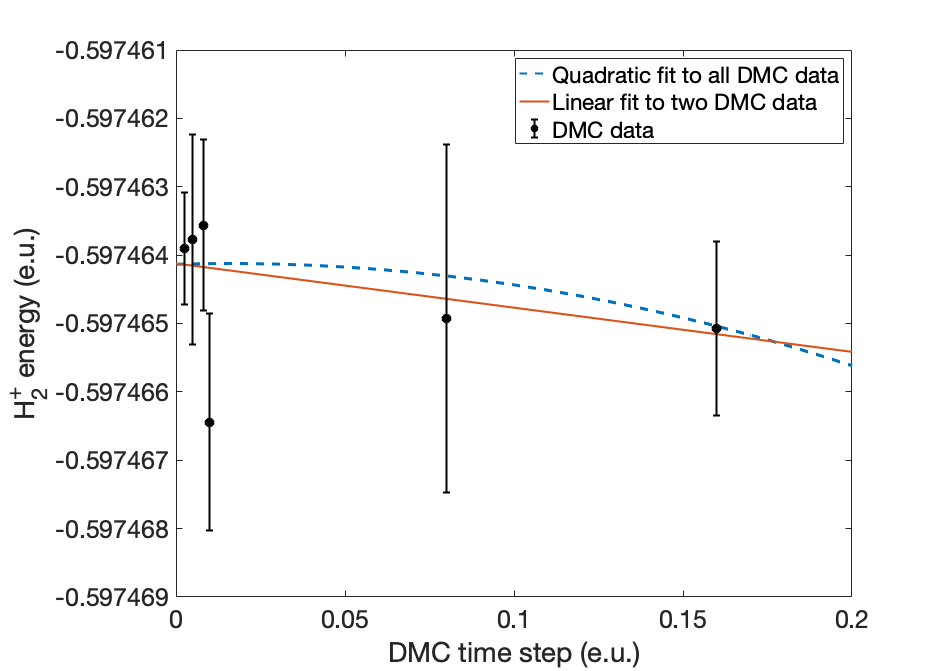}
 \centering
 \caption{DMC total energy against time step for an H$_2^+$ ion (a
   positive trion with a very small electron-hole mass ratio of
   $\sigma \approx 0.0005446$). A quadratic fit to the DMC energies at
   time steps of $0.0025$, $0.005$, $0.008$, $0.01$, $0.08$, and
   $0.16$ e.u.\ is shown as a dashed blue line. A linear fit to the
   DMC energies at two small time steps of $0.0025$ and $0.01$
   e.u.\ is shown as a solid red line. The DMC energies extrapolated
   to zero time step using the quadratic and linear fits are
   $-0.5974637(4)$ and $-0.5974636(7)$ e.u.,
   respectively. \label{fig:app_trion_time_step}}
\end{figure}

\begin{figure}[!htbp]
 \includegraphics[clip,width=1\linewidth]{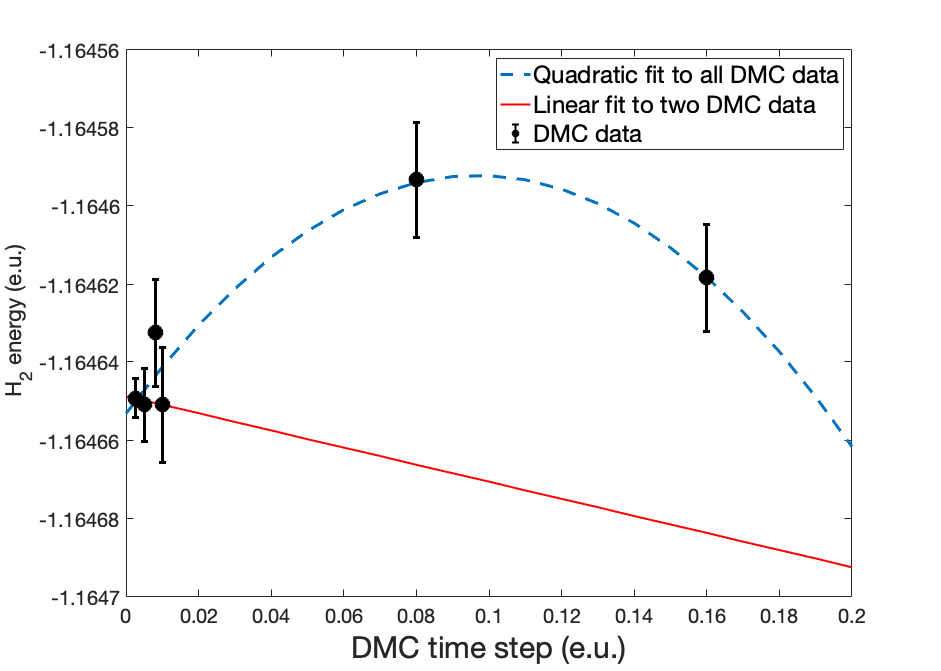}
 \centering
 \caption{As Fig.\ \ref{fig:app_trion_time_step}, but for an H$_2$
   molecule. The DMC energies extrapolated to zero time step using
   quadratic and linear fits are $-1.164653(5)$ and $-1.164649(8)$
   e.u., respectively. \label{fig:app_biexciton_time_step}}
\end{figure}

Figures \ref{fig:app_trion_time_step} and
\ref{fig:app_biexciton_time_step} show extrapolation of DMC energies
to zero time step for the H$_2^+$ ion and the H$_2$ molecule,
respectively. In both figures the DMC energies at zero time step are
obtained from two fits: (i) a quadratic function fitted to six time
steps and (ii) a linear fit to two small time steps. The
zero-time-step results are the same to within the statistical error
bars.

\section{DMC total energies for trions and biexcitons \label{app:raw_data}}

The DMC energies of negative trions, positive trions, and biexcitons
are reported in Tables \ref{table:DMC_trion_neg},
\ref{table:DMC_trion_pos}, and \ref{table:DMC_Biexciton},
respectively.

\begin{table}[!htbp]
\centering
\caption{DMC total energies of negative trions against mass ratio
  $\sigma$.} \label{table:DMC_trion_neg}

\begin{tabular}{lcc}
\hline \hline

\multirow{2}{*}{$\sigma$} & \multicolumn{2}{c}{Total energy (e.u.)} \\

& DMC (pres.\ work) & Prev.\ works \\

\hline

0 & $-0.527760(5)$ &  \begin{tabular}{@{}c@{}} $-0.5282$\footnote{Data
  taken from
  Ref.\ \onlinecite{Combescot_2017}. \label{fn:dist_Combescot}}
  \\ $-0.5275$\footnote{Data taken from
  Ref.\ \onlinecite{Usukura_1999}. \label{fn:dist_Usukura}} \end{tabular}
\\

0.05 & $-0.526318(4)$ & \\

0.10 & $-0.524994(6)$ & \\

0.20 & $-0.523792(4)$ & \\

0.32 & $-0.522915(7)$ & \\

0.40 & $-0.522653(6)$ & $-0.5222$\footref{fn:dist_Usukura} \\

0.50 & $-0.522550(6)$ & $-0.5246$\footref{fn:dist_Combescot} \\

0.64 & $-0.522737(8)$ & \\

0.70 & $-0.522881(7)$ & $-0.52275$\footref{fn:dist_Usukura} \\

0.80 & $-0.523191(8)$ & \\

0.90 & $-0.523599(8)$ & \\

1 & $-0.52401(1)$ & $-0.524$\footref{fn:dist_Usukura} \\

$\infty$ & $-0.60261(2)$ & \\

\hline \hline
\end{tabular}
\end{table}

\begin{table}[!htbp]
\centering
\caption{DMC total energies of positive trions against mass ratio
  $\sigma$. } \label{table:DMC_trion_pos}
\begin{tabular}{lc}
\hline \hline

$\sigma$ & DMC total energy (e.u.) \\ 

\hline

0 & $-0.60265(3)$ \\

0.05 & $-0.56283(3)$ \\

0.10 & $-0.551984(5)$ \\

0.20 & $-0.54094(1)$ \\

0.32 & $-0.534225(5)$ \\

0.40 & $-0.531452(5)$ \\

0.50 & $-0.529034(6)$ \\

0.64 & $-0.526793(7)$ \\

0.70 & $-0.52612(1)$ \\

0.80 & $-0.52521(2)$ \\

0.90 & $-0.524521(5)$ \\

1 & $-0.524002(5)$ \\

\hline \hline
\end{tabular}
\end{table}

\begin{table}[!htbp]
\centering
\caption{DMC total energy (in e.u.)\ of a biexciton X$_2$ against mass
  ratio $\sigma$.}\label{table:DMC_Biexciton}
\begin{tabular}{lc}
\hline \hline

$\sigma$ & DMC total energy (e.u.) \\

\hline

$0$ & $-1.17437(2)$ \\ 

$0.04$ & $-1.104619(7)$ \\ 

$ 0.05$ & $-1.098427(5)$ \\ 

$0.08$ & $-1.084599(4)$ \\ 

$0.1$ & $-1.07784(1)$ \\ 

$0.2$ & $-1.04545(2)$ \\

$0.4$ & $-1.04088(2)$ \\

$0.5$ & $-1.03715(2)$ \\

$0.62$ & $-1.034469(8$ \\ 

$0.725$ & $-1.033133(4)$ \\

$0.8$ & $-1.032556(4)$ \\ 

$0.92$ & $-1.03206(3)$ \\

$1$ & \begin{tabular}{@{}c@{}} $-1.03203(2)$
  \\ $-1.0321(1)$\footnote{Data taken from
  Ref.\ \onlinecite{Bressanini_1997}.} \end{tabular} \\

$1.6$ & $-1.0344(2)$ \\

$2.5$ & $-1.04086(1)$ \\

$4$ & $-1.05136(1)$ \\

$9$ & $-1.074611(4)$ \\

$20$ & $-1.098425(5)$ \\

$\infty$ & $-1.17440(2)$ \\ 

\hline \hline
\end{tabular}
\end{table}

\section{Born-Oppenheimer potential curves}

\subsection{DMC energies and fits \label{app:dmc_bo_data}}

DMC energies against hole-hole separation for positive trions and
biexcitons are reported in Tables \ref{table:trion_BO} and
\ref{table:DMC_biexciton_BO}, respectively.  The parameters in the
polynomials [Eqs.\ (\ref{eq:BO_fit}) and (\ref{eq:biex_BO})] fitted to
the DMC BO data are reported in Tables \ref{table:trion_BO_parameters}
and \ref{table:biex_BO_fitted_params}, respectively.

\begin{table}[!htbp]
\centering
\caption{DMC total energy of a positive trion in the heavy hole limit
  ($\sigma=0$), i.e., the BO potential energy $U_\text{hh}$. In each
  calculation, the two holes are fixed with separation
  $r_\text{hh}$. \label{table:trion_BO}}
\begin{tabular}{lcc}
\hline \hline

\multirow{2}{*}{$r_\text{hh}$ (e.u.)} &
\multicolumn{2}{c}{$U_\text{hh}$ (e.u.)} \\

& DMC (pres.\ wk.) & VMC\footnote{Data taken from
Ref.\ \onlinecite{Alexander_2005}.} \\

\hline

 $1.4$ & $-0.56998(2)$ & $-0.569983491(6)$ \\ 

$1.5$ & $-0.58233(2)$ & $-0.582323174(5)$ \\ 

$1.6$ & $-0.59092(2)$ & $-0.590937199(5)$ \\ 

$1.7$ & $-0.59672(2)$ & $-0.596696250(4)$ \\ 

$1.8$ & $-0.60024(2)$ & $-0.600253616(4)$ \\ 

$1.9$ & $-0.60209(2)$ & $-0.602105768(3)$ \\ 

$2$ & $-0.60265(2)$ & $-0.602634202(3)$ \\ 

$2.1$ & $-0.60215(2)$ & $-0.602134935(3)$ \\ 

$2.2$ & $-0.60082(2)$ & $-0.600839617(3)$ \\ 

$2.3$ & $-0.59891(2)$ & $-0.598930879(3)$ \\ 

$2.4$ & $-0.59652(2)$ & $-0.596553632(3)$ \\ 

$2.5$ & $-0.59381(2)$ & $-0.593823505(2)$ \\ 

$2.6$ & $-0.59084(2)$ & $-0.590833192(2)$ \\

$2.7$ & $-0.58765(2)$ & $-$ \\ 

$3.0$ & $-0.57756(3)$ & $-0.577562861(2)$ \\ 

$3.2$ & $-0.57069(2)$ & $-$ \\ 

$3.5$ & $-0.56092(8) $ & $-$ \\ 

\hline \hline
\end{tabular}
\end{table}

\begin{table}[!htbp]
\centering
\caption{DMC total energy of a biexciton in the heavy hole limit
  ($\sigma=0$), i.e., the BO potential energy $U_\text{hh}$. In each
  calculation, the two holes are fixed with separation
  $r_\text{hh}$. \label{table:DMC_biexciton_BO}}
\begin{tabular}{lcc}
\hline \hline

\multirow{2}{*}{$r_\text{hh}$ (e.u.)} &
\multicolumn{2}{c}{$U_\text{hh}$ (e.u.)} \\

& DMC (pres.\ wk.) & Prev.\ wk.\footnote{Data taken from
Ref.\ \onlinecite{Alexander_2004}.} \\

\hline

$0.2$ & $2.1976(2)$ & $2.197807(4)$ \\

$0.8$ & $-1.02007(2)$ & $-1.020056(1)$ \\

$0.9$ & $-1.08362(2)$ \\

$1$ & $-1.12449(2) $ & $-1.124539(2)$ \\

$1.1$ & $-1.15005(2)$ \\

$1.2$ & $-1.16488(2)$ \\

$1.3$ & $-1.17231(2)$ \\

$1.4$ & $-1.17445(3)$ & $-1.174475(3)$ \\

$1.5$ & $-1.17280(3)$ \\

$1.6$ & $-1.16860(3)$ \\

$1.7$ & $-1.16244(3)$ \\

$1.8$ & $-1.15503(3)$ & $-1.1550699(2)$ \\

$1.9$ & $-1.14683(3)$ & \\

$2 $ & $-1.13818(3)$ & \\

$2.1$ & $-1.12917(3)$ & \\

 $2.5$ & $-1.0881(10)$ & \\

\hline \hline
\end{tabular}
\end{table}

\begin{table}[!htbp]
\centering
\caption{Coefficients of the fitting function Eq.\ (\ref{eq:BO_fit})
  for the BO potential of a positive trion, determined by fitting to
  the data in Table \ref{table:trion_BO} in the range 1.4 a.u.${} \leq
  r_\text{hh} \leq 3.2$ a.u. \label{table:trion_BO_parameters}}
\begin{tabular}{lc}
\hline \hline

Parameter & Value (e.u.) \\

\hline

$p_0$ & $1.0427824014296136$ \\

$p_1$ & $-3.3397706586457314$ \\

$p_2$ & $ 2.8588375211577741$ \\

$p_3$ & $-1.3365671624439441$ \\

$p_4$ & $0.36129754841667888$ \\

$p_5$ & $-0.053146452736184853$ \\

$p_6$ & $0.0033007257815276473$ \\

\hline \hline
\end{tabular}
\end{table}

\begin{table}[!htbp]
\caption{Coefficients of the fitting function Eq.\ (\ref{eq:biex_BO})
  for the BO potential of a biexciton, determined by fitting to the
  data in Table \ref{table:DMC_biexciton_BO} in the range 0.8 a.u.${}
  \leq r_\text{hh} \leq 1.9$ a.u. \label{table:biex_BO_fitted_params}}

\begin{tabular}{lcl}
\hline \hline

Parameter & Value (e.u.) \\

\hline

$p_0$ & $ 4.9372623192546792$ \\

$p_1$ & $-26.739727920240565$ \\

$p_2$ & $ 55.588815121898229$ \\

$p_3$ & $-70.359270025396583$ \\

$p_4$ & $57.928244487543246$ \\

$p_5$ & $-31.202873732362676 $ \\

$p_6$ & $10.619660811375310$ \\

$p_7$ & $-2.0737318548490848$ \\

$p_8$ & $0.17711891075798453$ \\

\hline \hline
\end{tabular}
\end{table}

\subsection{Comparison of polynomial and Morse potential fits to BO potential
\label{app:bo_fits}}

\begin{figure}[!htbp]
\centering
\includegraphics[clip,width=1\linewidth]{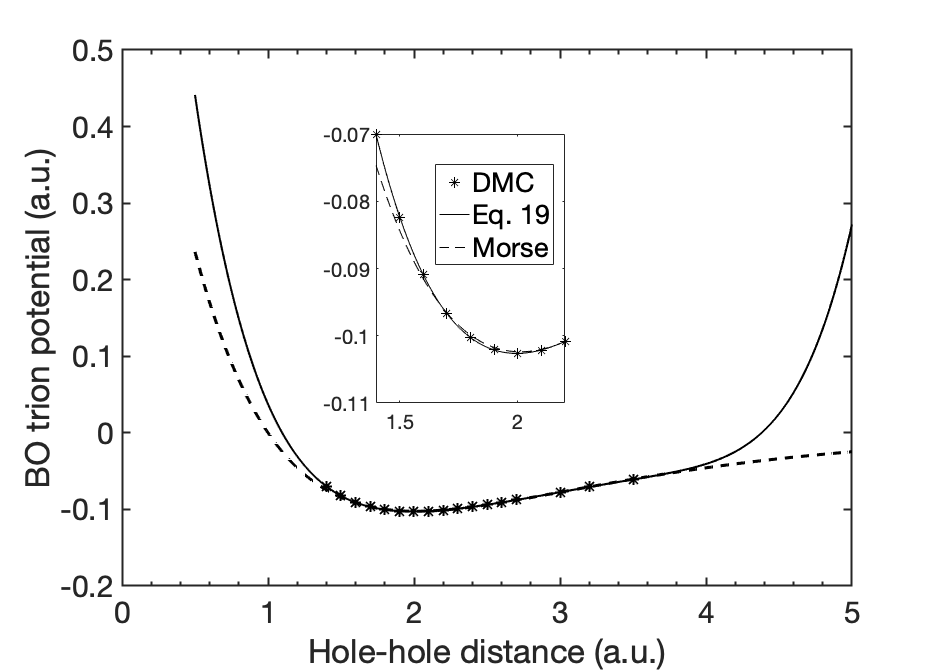}
\caption{Fits of a Morse potential [Eq.\ (\ref{eq:morse}), raw DMC
    data offset by $E_\text{X}$] and Eq.\ (\ref{eq:BO_fit}) to the DMC
  BO potential of a positive trion.  The Morse fitting parameters are
  $D_\text{eq}=0.102553(8)$ a.u., $R_\text{eq}=2.0181(1)$ a.u., and
  $a=0.7003(2)$ a.u.  The SSE and RMSE are $6.451 \times 10^{-6}$
  a.u.\ and $0.0006788$ a.u., respectively. The value of $R_\text{eq}$
  is slightly larger than the value predicted by fitting
  Eq.\ (\ref{eq:BO_fit}). However, the Morse potential shows much more
  reasonable behavior at small and large hole-hole separations. The
  fitted function given by Eq.\ (\ref{eq:BO_fit}) increases
  unphysically at large separations beyond $2.6$ a.u. However,
  Eq.\ (\ref{eq:BO_fit}) fits the DMC data very well in the vicinity
  of the equilibrium point, as seen in the inset; consequently
  spectroscopic constants predicted by Eq.\ (\ref{eq:BO_fit}), shown
  in Table \ref{table:trion_rdf_spec}, are more accurate than those
  predicted by the Morse potential, shown in Table
  \ref{table:Morse_trion_spec}. \label{fig:Morse_trion}}
\end{figure}

\begin{figure}[!htbp]
\centering
\includegraphics[clip,width=1\linewidth]{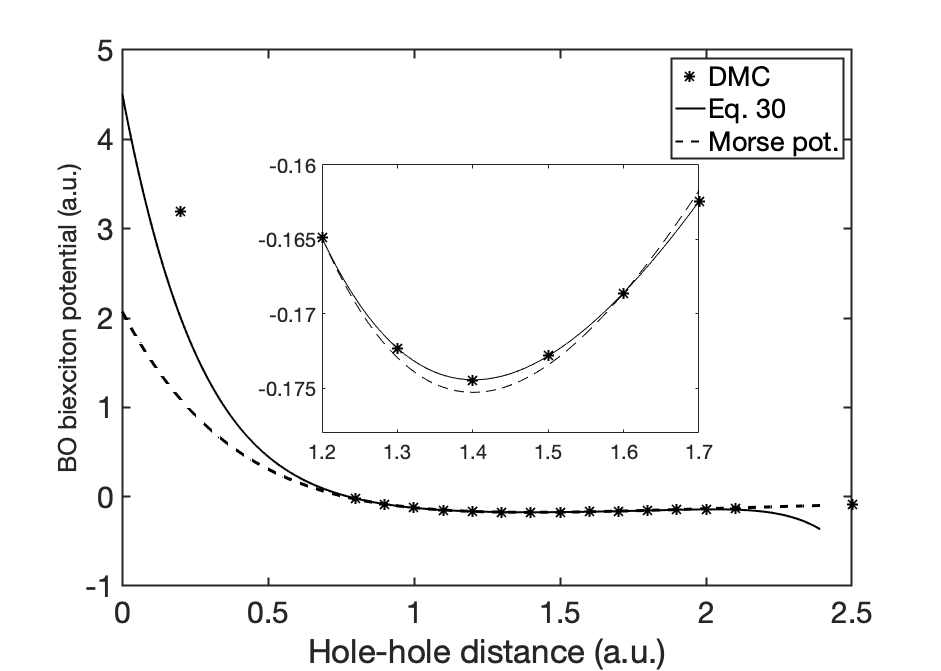}
\caption{Fits of a Morse potential [Eq.\ (\ref{eq:morse}), raw DMC
    data offset by $2E_\text{X}$] and Eq.\ (\ref{eq:biex_BO}) to the
  DMC BO potential of a biexciton. To obtain a better SSE and RMSE
  using the Morse model, we only included separations in the range 1
  a.u.${} \leq r_\text{hh} \leq {}$1.8 a.u.\ in the fit.  The Morse
  fitting parameters are $D_\text{eq}=0.17459(1)$ a.u.,
  $R_\text{eq}=1.4054(1)$ a.u., and $a=1.0402(6)$ a.u.\ The SSE and
  RMSE are $1.018 \times 10^{-5}$ a.u.\ and RMSE${}=0.001128$ a.u.,
  respectively. The $R_\text{eq}$ is close to the value obtained by
  fitting Eq.\ (\ref{eq:biex_BO}). In the large separation limit,
  contrary to the Morse model, Eq.\ (\ref{eq:biex_BO}) becomes
  unphysical. In the vicinity of the equilibrium point,
  Eq.\ (\ref{eq:biex_BO}) fits the DMC data better and produces more
  accurate spectroscopic constants, as seen by comparing Tables
  \ref{table:biex_spect} and \ref{table:Morse_biexciton_spec}.}
\label{fig:Morse_biexciton}
\end{figure}

A Morse interatomic potential is of the form
\begin{equation} U_\text{Morse}(r) = D_\text{eq} \left[ e^{-2a(r-R_\text{eq})}
-2e^{-a(r-R_\text{eq})} \right], \label{eq:morse} \end{equation} where
the equilibrium separation $R_\text{eq}$, well depth $D_\text{eq}$,
and $a$ are fitting parameters \cite{Morse_1929}.  By construction,
the Morse potential goes to zero at large separation, whereas the BO
potential goes to $E_\text{X}$ in a trion and to $2E_\text{X}$ in a
biexciton, where $E_\text{X}=-1/2$ e.u.\ is the ground-state energy of
a single exciton; hence we fit $U_\text{Morse}(r)$ to the
$U_\text{DMC}(r)-E_\text{X}$ raw data for trion and to the
$U_\text{DMC}(r)-2E_\text{X}$ for the biexciton.  Plots of
DMC-calculated points on the BO potential energy curves, together with
fitted polynomials and Morse potentials, are shown in
Figs.\ \ref{fig:Morse_trion} and \ref{fig:Morse_biexciton} for
dihydrogen cations and molecules, respectively.  The corresponding
spectroscopic constants are shown in Tables
\ref{table:Morse_trion_spec} and \ref{table:Morse_biexciton_spec}.

\begin{table}[!htbp]
\centering
\caption{Spectroscopic constants obtained from the Morse potential
  fitted to the DMC BO potential of the positive trion. The fitted
  parameters are listed in the caption of
  Fig.\ \ref{fig:Morse_trion}. \label{table:Morse_trion_spec}}
\begin{tabular}{lcccc}
\hline \hline

Cation & $\omega_\text{e}$ (cm$^{-1}$) & $\omega_\text{e}x_\text{e}$
(cm$^{-1}$) & $\alpha_\text{e}$ (cm$^{-1}$) & $B_\text{e}$ (cm$^{-1}$)
\\

\hline

H$_2^+$ & $2297.4(8)$ & $58.63(4)$ & $0.9298(7)$ & $29.348(4)$ \\

D$_2^+$ & $1624.9(6)$ & $29.33(2)$ & $ 0.3290(3)$ & $14.681(2)$ \\

T$_2^+$ & $1327.8(5)$ & $19.58(1)$ & $0.1795(1)$ & $9.803(1)$ \\

\hline \hline
\end{tabular}
\end{table}

\begin{table}[!htbp]
\centering
\caption{Spectroscopic constants obtained from the Morse potential
  fitted to the DMC BO potential of a biexciton. The fitted parameters
  are listed in the caption of
  Fig.\ \ref{fig:Morse_biexciton}. \label{table:Morse_biexciton_spec}}
\begin{tabular}{lcccc}
\hline \hline

Molecule & $\omega_\text{e}$ (cm$^{-1}$) & $\omega_\text{e}x_\text{e}$
(cm$^{-1}$) & $\alpha_\text{e}$ (cm$^{-1}$) & $B_\text{e}$ (cm$^{-1}$)
\\

\hline

H$_2$ & $4453(3)$ & $129.3(1)$ & $2.280(3)$ & $60.51(1)$ \\

D$_2$ & $3149(2)$ & $64.70(7)$ & $ 0.807(1)$ & $30.271(5)$ \\

T$_2$ & $2573(1)$ & $43.20(5)$ & $ 0.4401(5)$ & $20.213(4)$ \\

\hline \hline
\end{tabular}
\end{table}

\end{document}